\documentclass{article}
\usepackage[authoryear]{natbib}

\usepackage{arxiv}

\usepackage[utf8]{inputenc} 
\usepackage[T1]{fontenc}    
\usepackage{hyperref}       
\usepackage{url}            
\usepackage{booktabs}       
\usepackage{amsfonts}       
\usepackage{nicefrac}       
\usepackage{microtype}      
\usepackage{lipsum}
\usepackage{amsmath,amsfonts,amssymb,amscd}
\usepackage{xcolor}
\usepackage{hyperref}
\usepackage{graphicx}
\graphicspath{ {./images/} }
\usepackage{lineno}
\usepackage{array}
\usepackage{caption}
\usepackage{subcaption}
\usepackage{xcolor}

\title{The 4D Human Embryonic Brain Atlas: spatiotemporal atlas generation for rapid anatomical changes}

\author{
 Wietske A.P. Bastiaansen \\
 Biomedical Imaging Group Rotterdam, Department of Radiology and Nuclear Medicine, Erasmus MC, \\ Rotterdam, The Netherlands\\
Department of Obstetrics and Gynaecology, Erasmus MC, Rotterdam, The Netherlands\\
  \texttt{w.bastiaansen@erasmusmc.nl} \\
  Erasmus MC, P.O. Box 2040, 3000 CA Rotterdam, The Netherlands\\
  Corresponding author
   \And
 Melek Rousian \\
  Department of Obstetrics and Gynaecology, Erasmus MC, Rotterdam, The Netherlands\\
  \And
   Anton H.J. Koning  \\
Department of Pathology, Erasmus MC, Rotterdam, The Netherlands\\
  \And
   Wiro J. Niessen  \\
University Medical Center Groningen, The Netherlands
  \And
  Bernadette S. de Bakker \\
Amsterdam UMC location University of Amsterdam, Department of Obstetrics and Gynecology, The Netherlands\\
Amsterdam Reproduction and Development research institute, Amsterdam, The Netherlands\\
Department of Pediatric Surgery, Erasmus MC, Rotterdam, The Netherlands
\And
Régine P.M. Steegers-Theunissen  \\
Department of Obstetrics and Gynaecology, Erasmus MC, Rotterdam, The Netherlands\\
  \And
   Stefan Klein  \\
Biomedical Imaging Group Rotterdam, Department of Radiology and Nuclear Medicine, Erasmus MC, \\Rotterdam, The Netherlands\\
}

\begin{document}
\maketitle

\newpage

\begin{abstract}
Early brain development is crucial for lifelong neurodevelopmental health. However, current clinical practice offers limited knowledge of normal embryonic brain anatomy on ultrasound, despite the brain undergoing rapid changes within the time-span of days. To provide detailed insights into normal brain development and identify deviations, we created the 4D Human Embryonic Brain Atlas using a deep learning-based approach for groupwise registration and spatiotemporal atlas generation. Our method introduced a time-dependent initial atlas and penalized deviations from it, ensuring age-specific anatomy was maintained throughout rapid development. The atlas was generated and validated using 831 3D ultrasound images from 402 subjects in the Rotterdam Periconceptional Cohort, acquired between gestational weeks 8 and 12. We evaluated the effectiveness of our approach with an ablation study, which demonstrated that incorporating a time-dependent initial atlas and penalization produced anatomically accurate results. In contrast, omitting these adaptations led to an anatomically incorrect atlas. Visual comparisons with an existing ex-vivo embryo atlas further confirmed the anatomical accuracy of our atlas. In conclusion, the proposed method successfully captures the rapid anotomical development of the embryonic brain. The resulting 4D Human Embryonic Brain Atlas provides a unique insights into this crucial early life period and holds the potential for improving the detection, prevention, and treatment of prenatal neurodevelopmental disorders.
\end{abstract}


\newpage
\section{Introduction}
Prenatal brain development is of crucial importance for the neurodevelopmental health of the developing embryo, fetus, and offspring. During the first trimester, brain development is monitored by visual inspection of standard planes using two-dimensional ultrasound \citep{Paladini2020, Paladini2021}. However, most congenital brain anomalies are missed during the first trimester, since sonographers often lack detailed knowledge of the appearance of the embryonic brain on ultrasound \citep{Volpe2021,Buijtendijk2024}. More subtle differences in early brain morphology caused by maternal conditions such as age and obesity, and lifestyle behavior, such as nutrition, smoking, and alcohol use, may be even harder or impossible to visually distinguish. However, these differences are relevant: for example, maternal obesity is associated with a decreased trans-cerebellar diameter during the first trimester \citep{Koning2017}, smaller overall fetal size during the first trimester \citep{Thagaard2018, Vanzundert2022}, fetal complications \citep{Guelinckx2008} and miscarriages \citep{Metwally2008}. 

To study the normal and abnormal variations in early brain morphology in more detail, several authors have proposed a spatiotemporal atlas of the embryonic and fetal brain. A spatiotemporal atlas consists of a set of brain templates, based on imaging data from a population, and can be defined as an average model from a geometrical (shape) and iconic (image intensity) point of view \citep{Oishi2019,Legouhy2022}. Table \ref{tab:relatedwork} shows an overview of population-based fetal brain atlases. All these atlases concern the second trimester and later and most are based on magnetic resonance imaging (MRI). Only the atlases by \citet{Namburete2018, Namburete2023} are based on ultrasound, but these do not include the first trimester. Not mentioned in Table \ref{tab:relatedwork}, is the 3D Embryo Atlas by \citet{Bakker2016}. This atlas consists of interactive 3D digital reconstructions based on histologically sectioned human embryos of the Carnegie collection and is therefore not a population-based atlas. Hence, we conclude that there is a lack of a population-based ultrasound atlas covering the first trimester. Moreover, except for the atlas created by \citet{Pei2021}, all atlases have a temporal resolution of one week. Significant anatomical changes, as described by the Carnegie stages of embryonic development, appear in a time-span of days \citep{Orahilly2008}. Therefore, a template covering every gestational day would capture the rapidly changing anatomy of the brain (e.g., the appearance of the choroid plexus, changes in neck curvature and development of the ventricles) more completely.

For all studies in Table \ref{tab:relatedwork}, the atlas construction procedure heavily relied on image registration techniques. Two main approaches can be distinguished, based on either multiple pairwise registrations or on a single groupwise registration. In pairwise registration, all images in the population are registered to each other and the atlas is obtained by taking the mean of all these mappings \citep{Seghers2004}. In groupwise image registration, all images are registered simultaneously and the atlas is implicitly defined by constraining the sum of all deformations from the atlas to each subject to be zero \citep{Balci2007,Bhatia2004} (groupwise constraint). Both these approaches lead to an atlas that is unbiased towards subjects in the population. \citet{Dalca2019} created a deep learning-based approach for groupwise image registration and atlas generation, which consists of two networks: one to learn how to register images to the atlas, and one to learn how to generate the atlas based on additional information such as age and sex of the subject. Shorter computation time during inference, when registering images of new subjects to the atlas, makes this approach favorable over non-learning based approaches. From Table \ref{tab:relatedwork}, \citet{Pei2021} used a learning-based approach. Their work is an extension of the work by \citet{Dalca2019}, where they used available segmentations of brain structures to guide the generation of the atlas. 

We extended the work by \citet{Dalca2019}, which was evaluated for Alzheimer’s disease characterized by relatively subtle anatomical changes over time. We extend their approach by introducing a time-dependent initial atlas and penalizing deviations from it, ensuring that rapid anatomical changes occurring over a span of days in first trimester brain development are preserved. The main contributions of research presented in this paper are:
\begin{enumerate}
\item A deep learning–based approach for spatiotemporal atlas creation\footnote{Preliminary results were published in \citet{Bastiaansen2022wbir}. There, we built upon the work by \citet{Dalca2019}, but instead of jointly registering images to and generating the atlas, the images were registered to the atlas and the atlas was updated iteratively.} adapted to account for rapid anatomical changes, validated through an ablation study showing that incorporating a time-dependent initial atlas and penalization of deviations improve anatomical accuracy compared to models without these adaptations.
\item The 4D Human Embryonic Brain Atlas, the first spatiotemporal brain atlas of the first trimester based on ultrasound imaging, which provides a framework for analyzing the influence of parental and lifestyle factors on early brain development, supports early detection of congenital anomalies, and serves as a clinically valuable visualization tool to enhance parental understanding and engagement.
\item An evaluation of the anatomical correctness of the resulting atlas by comparing its volume to a reference head volume (HV) curve established in previous research \citep{Koning2016}, and through visual comparison to the 3D Embryo Atlas by \citet{Bakker2016} and the Fetal Brain Atlas by \citet{Namburete2023}.
\item A study assessing whether the atlas can capture known morphological differences associated with maternal overweight and obesity during early pregnancy using Voxel-Based Morphometry \citep{Ashburner2000}.
\end{enumerate}

\begin{table}[h!]
 \caption{Literature overview of existing population-based fetal brain atlases. MRI = magnetic resonance imaging, US = ultrasound, dMRI = diffusion magnetic resonance imaging.}
  \centering
  \setlength\extrarowheight{5pt}
  \begin{tabular}{>{\raggedright}m{0.13\textwidth}>{\centering}m{0.07\textwidth}>{\centering}m{0.08\textwidth}>{\centering}m{0.15\textwidth}>{\centering}m{0.15\textwidth}>{\centering}m{0.1\textwidth}>{\centering\arraybackslash}m{0.1\textwidth}}
    \toprule
     & \textbf{Age} (weeks) & \textbf{Modality} & \textbf{Acquisition} & \textbf{Registration} & \textbf{Temporal resolution} & \textbf{Number of subjects} \\
     \toprule
     \citet{Habas2010} & 20 -- 24 & MRI & During pregnancy & Groupwise& Weeks & 20 \\
     \hline
     \citet{Kuklisova2011}& 29 -- 44 & MRI & After premature birth &Pairwise&Weeks& 142 \\
     \hline
     \citet{Serag2012}& 26 -- 44 & MRI & After premature birth &Pairwise &Weeks &204\\
     \hline
     \citet{Zhan2013}& 15 -- 22 & MRI & Postmortem &Pairwise &Weeks& 34 \\
     \hline
     \citet{Gholipour2007} & 19 -- 39 & MRI & During pregnancy &Pairwise &Weeks&81 \\
     \hline
     \citet{Namburete2018} & 23 & US & During pregnancy &Groupwise &Weeks & 39 \\
     \hline
     \citet{Khan2019} & 21 -- 39 & MRI & During pregnancy &Pairwise &Weeks& 67 \\
     \hline
     \citet{Uus2021} & 23 -- 44 & dMRI & After (pre)mature birth &Pairwise& Weeks &210\\
     \hline
     \citet{Pei2021} & 22 -- 31 & MRI &During pregnancy & Groupwise&  Days & 82 \\
     \hline
     \citet{Wu2021} & 20 -- 39 & MRI & During pregnancy &Groupwise &Weeks &735 \\
     \hline
     \citet{Chen2022deciphering} & 24 -- 38 & dMRI & During pregnancy&Pairwise & Weeks& 89 \\
     \hline
     \citet{Xu2022} & 23 -- 28 & MRI & During pregnancy & Pairwise & Weeks & 90\\
     \hline
    \citet{Namburete2023} & 14 -- 30 & US & During pregnancy & Groupwise & Weeks & 899\\
    \bottomrule
    \textbf{Our contribution} & \textbf{8 -- 12} & \textbf{US} & \textbf{During pregnancy} & \textbf{Groupwise} & \textbf{Days} & \textbf{308} \\
    \bottomrule
  \end{tabular}
  \label{tab:relatedwork}
\end{table}
\clearpage
\section{Method}

\begin{figure}[b!]
  \centering
  \includegraphics[width=\textwidth]{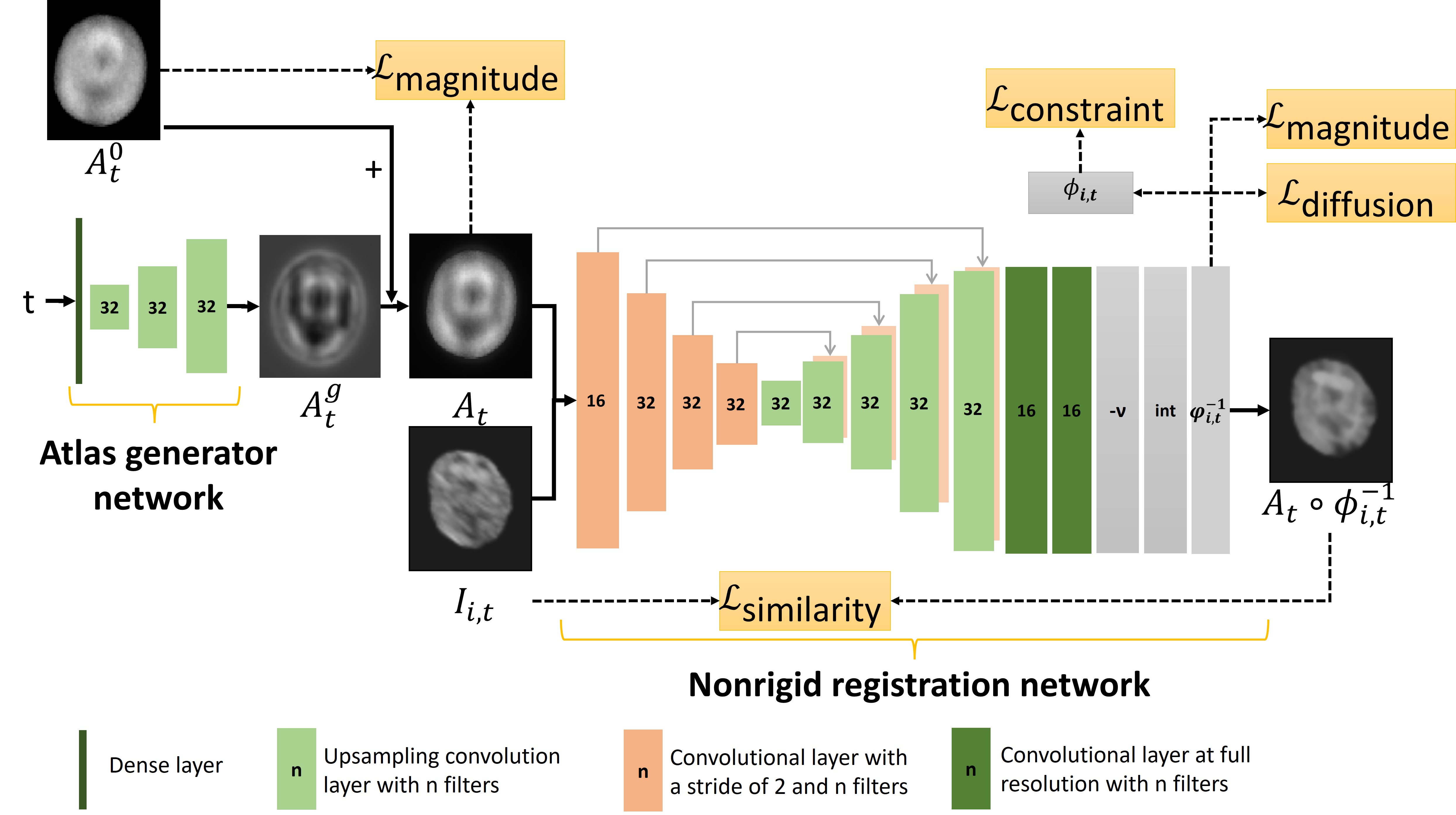}
  \caption{Overview of the proposed method and network architectures for spatiotemporal atlas generation. The \textit{atlas generator network} takes as input the gestational age $t$ (in days, normalized between 0 and 1) and the initial atlas $A_t^0$, and outputs the generated atlas $A_t = A_t^0 + A_t^g$. It consists of a dense layer followed by three upsampling convolutional layers with 32 filters, with the number of upsampling layers chosen such that the final resolution matches that of the input images $I_{i,t}$. A final convolutional layer at full resolution produces $A_t^g$. The \textit{nonrigid registration network} follows the VoxelMorph architecture \citep{Balakrishnan2019} and includes an encoder with one convolutional layer of 16 filters and three layers with 32 filters (stride 2), a decoder with five upsampling convolutional layers with 32 filters, skip connections, and two convolutional layers with 16 filters at full resolution. Diffeomorphic deformations are obtained using a stationary velocity field representation, where the deformation field $\phi_{i,t}$ is computed by integrating the velocity field $\nu_{i,t}$, and the inverse deformation $\phi_{i,t}^{-1}$ by integrating its negation $-\nu_{i,t}$ \citep{Ashburner2007}.}.
  \label{fig:fig_overview_method}
\end{figure}

The proposed method generates a spatiotemporal atlas $A_t$ generated from 3D images $I_{i,t}$, from subject $i$ imaged at time $t$, being the gestational age (GA) in days. The method consists of two networks that are trained simultaneously \citep{Dalca2019}. Figure \ref{fig:fig_overview_method} gives an overview. The first network generates the atlas and takes as input an initial atlas $A_t^0$ and the time $t$ and outputs $A_t=A_t^0+A_t^g$, where $A_t^g$ is the generated deviation from the initial atlas. The second network is the nonrigid registration network, which takes as input the atlas $A_t$ and an image $I_{i,t}$ and gives as output the deformation field $\phi_{i,t}$ and its inverse. $\phi_{i,t}$ registers each image to the atlas, and can therefore be used to describe morphological variations within the population. In the remainder of this section, all components shown in the figure are explained in more detail.

\subsection{Atlas $A_t$ and images $I_{i,t}$}
The raw images $\overline{I_{i,t}}$ are defined on voxel grid $\Omega_{\overline{I_{i,t}}}$ with variable isotropic voxel spacing $s_{I_{i,t}}$ in $mm$ and $n_I$ the maximum number of voxels over all images $\overline{I_{i,t}}$. The atlas $A_t$ is defined on voxel grid $\Omega_{A_t}$ with size $n_{A_t} \times n_{A_t}  \times n_{A_t} $ with isotropic time-dependent voxel spacing $s_{A_t}$ in $mm$. The images $I_{i,t}$ are created by rigidly aligning $\overline{I_{i,t}}$ to a standard orientation, re-sampling to voxel size $s_{A_t}$ and cropping to a voxel grid of size $n_{A_t} \times n_{A_t} \times n_{A_t} $.  After resampling all images to a time-dependent voxel spacing $s_{A_t}$, the brain fills a similar field of view over time, which simplifies the training of the non-rigid registration network. Furthermore, we assume that the embryonic head is segmented in the images. We call $S_{I_{i,t}}$ the segmentation of the head in image $I_{i,t}$, and $S_{A_{t}}$ the segmentation of the head in the atlas at time $t$.

\subsection{The time-dependent initial atlas $A_t^0$}
The network generates the atlas by learning the deviation $A_t^g$ from the initial atlas, which is an input of the network together with the time $t$. We create a time-dependent initial atlas, instead of using a single initial atlas as proposed by \citet{Dalca2019}, to account for the rapid morphological changes (e.g., the appearance of the choroid plexus, change of curvature of the embryonic body). The initial atlas $A_t^0$ is computed as the voxelwise median over all images $I_{i,t}$ in a time window $\tau \in[t-\delta, t+\delta]$. The median was chosen over the mean to produce a sharper initial atlas, as previously shown \citep{Bastiaansen2022wbir}. $\delta$ is a hyperparameter of the network.

\subsection{Network architecture}
We used the default network architecture as described by \citet{Dalca2019}. The atlas generator network takes as input the GA $t$ in days normalized between 0 and 1, followed by a dense layer and three upsampling convolution layers with 32 filters. The number of upsampling layers is chosen such that the final layer has the resolution of the images $I_{i,t}$. The upsampling layers are followed by a convolutional layer on full resolution that results in $A_t^g$.

For the nonrigid registration network, the default architecture of VoxelMorph was used \citep{Balakrishnan2019}. The network consists of an encoder,  with convolutional layers with a stride of 2, a decoder with upsampling convolutional layers, skip connections, followed by two convolutional layers at full resolution with 16 filters. The encoder consists of one layer with 16 filters and is followed by 3 layers with 32 filters. The decoder consists of 5 upsampling layers with 32 filters. 

To ensure invertibility of the deformation $\phi_{i,t}^{-1}$, we use diffeomorphic nonrigid deformations. Following \citet{Dalca2019}, we use a stationary velocity field representation, meaning that $\phi_{{i,j}}$ is obtained by integrating the velocity field $\nu_{i,j}$ and $\phi_{i,j}^{-1}$ is obtained by integrating the negated velocity field $-\nu_{i,j}$ \citep{Ashburner2007}. 

\subsection{Loss function}
The loss function is defined as follows:
\begin{align}
\begin{split}
    \mathcal{L} \left(A_t, A_t^g, I_{i,t},\phi_{i,t}, u_{i,t}\right) &= \mathcal{L}_{\text{similarity}}\left(A_t\circ \phi_{i,t}^{-1},I_{i,t}\right)+ \lambda_{\text{constraint}}\mathcal{L}_{\text{constraint}}\left(u_{i,t}\right)\\& + \lambda_{\text{deformation}}\left[\mathcal{L}_{\text{magnitude}}\left(u_{i,t}^{-1}\right)  + \mathcal{L}_{\text{diffusion}}\left(u_{i,t}^{-1}\right)\right]\\ &+ \lambda_{\text{atlas}} \mathcal{L}_{\text{magnitude}}(A_t^g)
\end{split}
\end{align}
with $u_{i,t}$ the spatial displacement for deformation field $\phi_{i,t}(x):= x + u_{i,t}(x)$. The first four terms were also used by \citet{Dalca2019}. Here, we introduce a fifth term that constrains the generated atlas by regulating deviations from the time-dependent initial atlas. In contrast, the original method by \citet{Dalca2019} uses a time-independent initial atlas, created by taking the voxel-wise mean of the training images. Due to the rapid anatomical development of the embryonic brain (e.g., the appearance of the choroid plexus), this time-independent approach results in an anatomically incorrect initial atlas, which can lead to a misrepresentation of embryonic brain development in the generated atlas, biased toward the inaccurate initial reference.

The first term $\mathcal{L}_{\text{similarity}}$ computes the similarity between the atlas and image. We use the local squared normalized cross-correlation using a 9x9x9 window to compute the average local intensity. 

The second term $\mathcal{L}_{\text{constraint}}$ by \citet{Dalca2019} constrains the average deformation to be the identity transform, which implies that $\frac{1}{N}\sum_{i,t} u_{i,t} \approx 0$ with $N$ the number of images used for training \citep{Balci2007, Bhatia2004}. $\mathcal{L}_{\text{constraint}}$ minimizes the running average over the last $\kappa$ deformation fields obtained at epoch $K$ during training with $u_{i,t}^{k}$ obtained at epoch $k$:
\begin{equation} \label{eq:constraint}\mathcal{L}_{\text{constraint}}\left(u_{i,t}\right) = \left \| \frac{1}{\kappa} \sum_{k=K-\kappa}^{K} u_{i,t}^k \right \|^2_2.\end{equation}

Here, we choose to let the constraint be satisfied approximately per gestational day. To this end, we sort the data during training based on GA and calculate the constraint over window $\kappa$, where $\kappa$ is the average number of images per gestational day in the training data.

To regularize both the obtained deformations $\phi_{i,t}$ and the atlas $A_t$ the following loss term is used:
\begin{equation}
\mathcal{L}_{\text{magnitude}}(U) = \|U  \|_2^2. \end{equation}
with a vector $U$. $\mathcal{L}_{\text{magnitude}}(u_{i,t}^{-1})$ penalizes large deformations and $\mathcal{L}_{\text{magnitude}}(A_t^g)$ penalizes large generated deviations from the initial template $A_t^0$.

Finally, the deformations are regularized by $\mathcal{L}_{\text{diffusion}}$, which penalizes local spatial variations in $\phi_{i,t}$ to promote smooth local deformations, and is defined as follows:
\begin{equation} \mathcal{L}_{\text{diffusion}} (u_{i,t}^{-1}) = \|\nabla u_{i,t}^{-1} \|_2^2. \end{equation}

\subsection{Implementation details}
The two networks were trained end-to-end using the ADAM optimizer with a learning rate of $10^{-6}$, chosen based on preliminary experiments. Training proceeded for 500 epochs with a batch size of one, with each image in the training set processed once per epoch to update the model weights. Validation loss was computed after each epoch to monitor convergence and evaluate performance. The training was carried out on an RTX 2080 Ti 11 GB GPU with an AMD EPYC 7452 CPU and 14 GB of RAM. The code used for these experiments is publicly available at \url{https://gitlab.com/radiology/prenatal-image-analysis/4D_human_embryonic_brain_atlas}
.

\subsection{Voxel-based morphometry (VBM)}
VBM can be used to study differences in morphology between subgroups of a population \citep{Ashburner2000}. To perform VBM, first all images $I_{i,t}$ are registered to the same reference, here to the atlas $A_t$, to obtain $\phi_{i,t}$. Next, a morphology descriptor based on $\phi_{i,t}$ is chosen and voxelwise maps are calculated. We choose the Jacobian determinant of $\phi_{i,t}$ in every voxel within the brain segmentation $S_{A_{t}}$. The Jacobian determinant in a voxel describes the volumetric change with respect to the atlas. Subsequently, the obtained descriptor maps are smoothed by a Gaussian filter and log-transformed to make the statistical distribution of the values symmetrical and normal. Here, the Jacobian determinants maps were smoothed using a Gaussian filter with $\sigma=2$. Finally, a statistical test is performed to assess whether a structural difference in morphology exists between the subgroups.

To perform VBM in a 4D reference space, the statistical comparison between subgroups can be performed at each gestational day. However, since we have limited data at some gestational days (less than 10 images), we choose to perform the statistical comparison within non-overlapping time windows of $2\delta+1$ days, for the same value of $\delta$ as used to create the initial atlas. Alternatively, the statistical comparison could be performed over the whole time period at once. We choose to perform the analysis per time window since the spatial correspondence throughout weeks 8 to 12 is limited by the appearance of structures (e.g., choroid plexus early week 9) and large changes in the curvature of the embryonic body. Within smaller time windows the spatial correspondence is maintained due to the time-dependent voxel sizes used and Gaussian smoothing of the Jacobian determinant.

\section{Data}
The Rotterdam Periconceptional Cohort (Predict study) is an ongoing hospital-based cohort study conducted at the Erasmus MC, University Medical Center Rotterdam, the Netherlands. The pilot study ran between 2009 and 2010 and the ongoing study started in 2010. This prospective cohort study focuses on the relationships between periconceptional maternal and paternal health and embryonic and fetal growth and development \citep{Rousian2021,Steegers2016}. 
3D ultrasound scans are acquired at multiple time points during the first trimester on a Voluson E8 (GE Healthcare, Austria) ultrasound machine with a 4.5 - 11.9 MHz trans-vaginal probe. Furthermore, data is collected regarding maternal lifestyle and dietary habits of the future parents (e.g., food frequency questionnaire), anthropometric measures (e.g., weight, height), and blood samples for biomarker analysis.

\subsection{In- and exclusion criteria}
Women were eligible to participate in the Rotterdam Periconceptional Cohort if they were at least 18 years of age, with an ongoing singleton pregnancy less than 10 weeks of GA. Our aim was to create an atlas describing normal brain development, therefore, we included only pregnancies with no adverse outcomes (miscarriage, termination of pregnancy, intra-uterine fetal death, congenital malformations, stillbirth, or postpartum death). Furthermore, we only included pregnancies from spontaneous conception, where the mother had a regular menstrual cycle. In this case, the GA can be calculated reliably according to the first day of the last menstrual period (LMP). Pregnancies with an unknown date of LMP were excluded. Within the included population the mean discrepancy between the GA calculation using LMP and crown-rump length (CRL) was -0.31 days with a standard deviation of 2.51 days, which indicates agreement between both methods and thus a strictly dated population. 
\newpage
\subsection{Atlas-generation, validation, and test dataset}
In total, we included 831 ultrasound images of 402 pregnancies of 392 subjects\footnote{10 women participated again in the study in a subsequent pregnancy} acquired between 56 and 90 days GA. The data was split into an atlas-generation, validation, and test set. This was done in such a way that approximately for every gestational day $80\%$ of the data was in the atlas-generation set, $10\%$ in the validation set, and $10\%$ in the test set. The atlas-generation set was used to train the network, the validation set was used to evaluate hyperparameter choices and perform the ablation study, and the test set was used to evaluate generalizability. Furthermore, the subjects fell into one of the following four maternal BMI categories: low: BMI < 19, normal: 19 $\leq$ BMI < 25, overweight: 25 $\leq$ BMI < 30, and obese: BMI $\geq$ 30. The BMI was calculated based on standardized weight and height measurements performed during the first visit. For $16\%$ of the subjects this was preconceptionally, for $78\%$ this was in gestational week 7, and for $5\%$ this was in gestational week 9. The data was split in such a way that the ratio between the four BMI categories was similar in all three datasets. 

\begin{figure}[t!]
     \centering
     \begin{subfigure}[b]{0.3\textwidth}
         \centering
         \includegraphics[width=\textwidth]{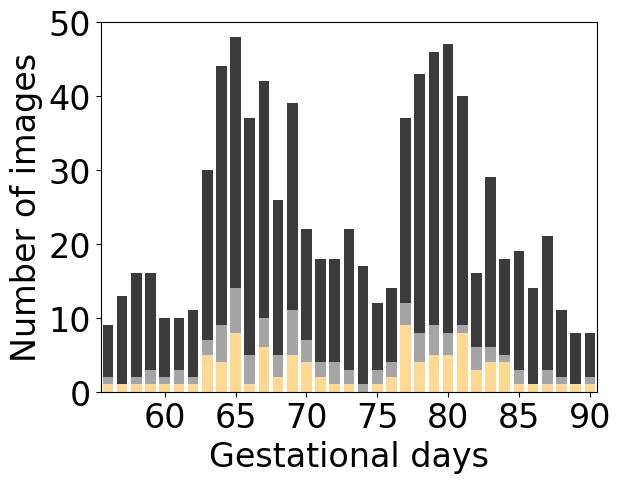}
         \caption{}
     \end{subfigure}
     \hfill
     \begin{subfigure}[b]{0.3\textwidth}
         \centering
         \includegraphics[width=\textwidth]{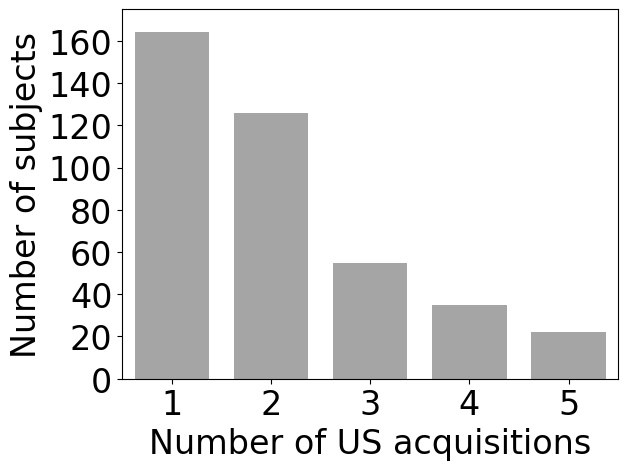}
         \caption{}
     \end{subfigure}
     \hfill
     \begin{subfigure}[b]{0.3\textwidth}
         \centering
         \includegraphics[width=\textwidth]{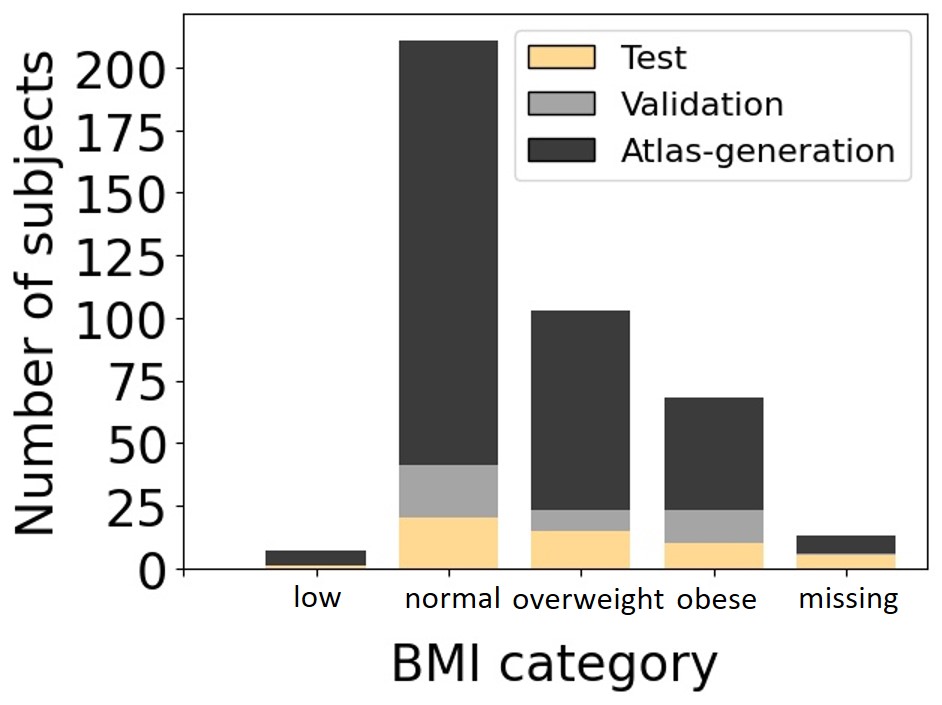}
         \caption{}
     \end{subfigure}
        \caption{Dataset characteristics. (a) The number of ultrasound scans available per gestational day for the atlas-generation set, validation set, test set and in total. (b) The number of subjects with $1,2, \ldots, 5$ ultrasound acquisitions. (c) The number of subjects per BMI category in the atlas-generation, validation and test set, where the categories are defined as follows: low: BMI < 19, normal: 19 $\leq$ BMI < 25, overweight:  25 $\leq$ BMI < 30, and obese: BMI $\geq$ 30.}
        \label{fig:fig_data_char}
\end{figure}

The atlas-generation set consists of 654 ultrasound scans from 308 subjects, the validation set consists of 80 ultrasound scans from 43 subjects, and the test set consists of 97 ultrasound scans from 51 subjects. Figure \ref{fig:fig_data_char} shows the number of available images per gestational day, the number of subjects with $1,2, \ldots, 5$ ultrasound acquisitions, and the number of images per BMI category. The number of available images per gestational day varies, which is due to the study protocol. In the pilot study, women received weekly ultrasound scans, while in the subsequent study, scans were performed only in gestational weeks 7, 9, and 11, as it was deemed sufficient to accurately model growth, while also relieving the burden on the participants \citep{Steegers2016,Uitert2013}.

\begin{figure}[b!]
  \centering
  \includegraphics[width=\textwidth]{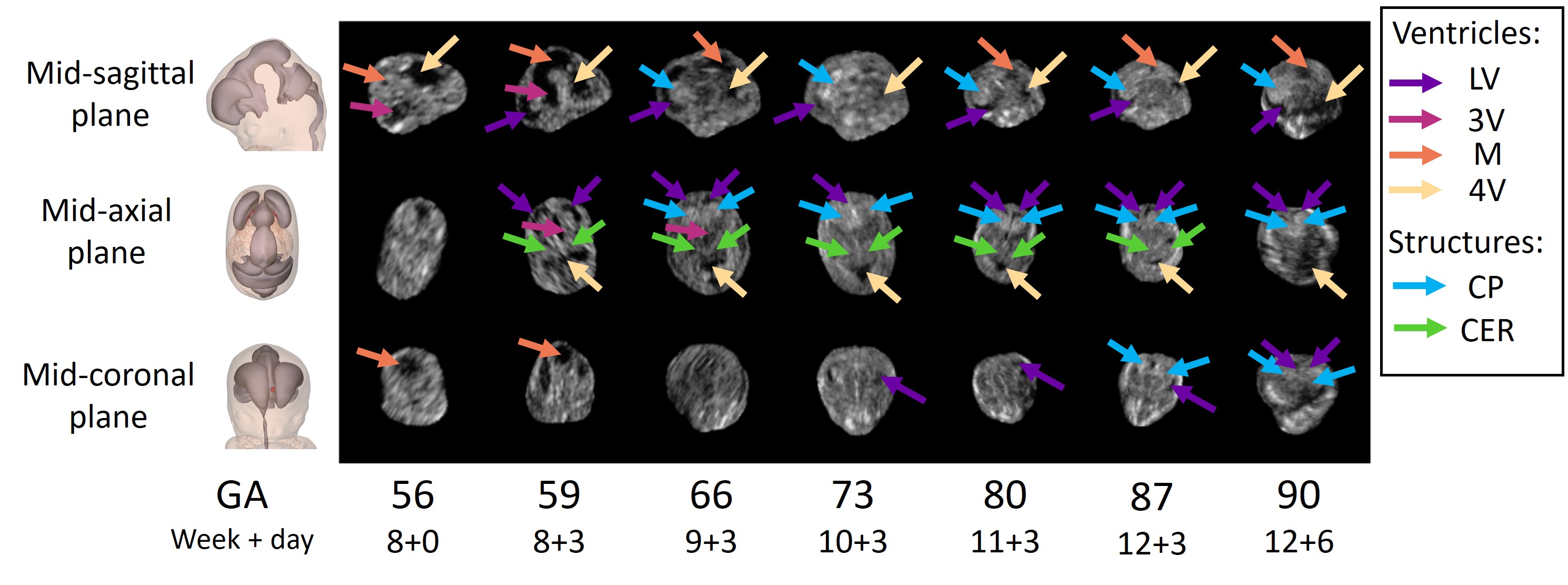}
  \caption{Example of mid-sagittal, mid-axial and mid-coronal plane at different gestational days after preprocessing of the imaging data. Brain structures are marked with arrows. LV = lateral ventricle, 3V = third ventricle, M = cavity of the mesencephalon, 4V = fourth ventricle, CP = choroid plexus, CER = cerebellum.}
  \label{fig:fig_overview_data}
\end{figure}

\subsection{Pre-processing of the images $I_{i,t}$}
The images $I_{i,t}$ were obtained from the raw images $\overline{I_{i,t}}$ by segmenting the embryonic head. Segmentation of the embryonic and fetal head was performed using nnU-net \citep{Isensee2021}, which was trained on segmentations created using an in-house developed virtual reality (VR) system \citep{Bastiaansen2025}. 

Segmentation was followed by rigid alignment of the head to a standard orientation. This was achieved by extracting the translation vector and rotation matrix from an affine transformation, which was obtained using our previously developed algorithm for multi-atlas registration of the embryo \citep{Bastiaansen2022}. The rotation matrix was extracted using the singular value decomposition. The previously developed algorithm aligned each image to a selected subset of atlases closest in gestational age, where the complete set of atlases consisted of ultrasound images acquired at a range of gestational ages, manually segmented, and spatially aligned to a predefined standard orientation.

Next, all images were resampled to voxel size $s_{A_t}$. The time-dependent voxel size $s_{A_t}$ was determined by linear regression of the median voxel size in $mm$ per gestational week of the raw images $\overline{I_{i,t}}$, resulting in: $S_{A_t} = -0.3606 + 0.0084t$. The voxel size varied from 0.11 x 0.11 x 0.11 mm at gestational day 56 to 0.40 x 0.40 x 0.40 mm at gestational day 90.

Finally, all images were center-cropped to a voxel grid of size $n_{A_t} \times n_{A_t} \times n_{A_t}$ and normalized to intensity values between 0 and 1. Here, we have chosen $n_{A_t}=128$. Figure \ref{fig:fig_overview_data} shows examples of preprocessed images at different days GA, visible brain structures are marked using arrows.

\section{Experiments}

\subsection{Evaluation metrics}
Registration accuracy was evaluated by calculating the Dice Similarity Coefficient (DSC) score between $S_{I_{i,t}}$ and $S_{A_{t}}\circ \phi_{i,t}^ {-1}$ \citep{Dice1945}. The segmentation of the atlas $S_{A_{t}}$ was obtained by including all voxels in $A_{t}$ with image intensity $\geq 0.1$, followed by a binary opening and closing to remove noise and include the hypo-intense brain ventricles. Both the opening and closing were done using a ball with a radius of 10 voxels. The regularity of $\phi_{i,t}$ was evaluated by calculating the percentage of voxels having a non-positive Jacobian determinant $\% |J|\leq 0$ within the mask $S_{A_t}$. For both the DSC and $\% |J|\leq 0$ the mean and standard deviation over the validation and/or test set were reported.

Volumetric growth of the atlas was evaluated by calculating the relative error of the head volume in $cm^3$ of the atlas $\text{V}(A_t)$ with respect to the head volume (HV) curve determined by \citet{Koning2016}: $\text{HV}_{\text{VR}}(t)=(-1.0947 + 0.0315t)^4$ with $t$ the gestational days. This curve was obtained by performing measurements of the HV using semi-automatic VR software. The relative $\text{HV}_{\text{error}}$ for every $t$ is calculated as follows:
\begin{equation}
    \text{HV}_{\text{error}}(A_t,t) = \frac{| \text{V}(A_t) - \text{HV}_{\text{VR}}(t)|}{\text{HV}_{\text{VR}}(t)} \cdot 100 \%
\end{equation}

To measure the sharpness of the atlas we used the metric proposed by \citet{Legouhy2022}, which is defined as follows:
\begin{equation}
    \text{sharpness}(A_t) = \text{mean}_{x \in S_{A_t}} \frac{\text{std}_j[A_t(x_j)]}{\text{mean}_j[A_t(x_j)]}
\end{equation}
where $j$ iterates over a $J^3$ volume around $x \in S_{A_t}$ with $J=5$. 

Finally, differences between generated atlases were evaluated by calculating the structural similarity index measure (SSIM) \citep{Wang2004}. The SSIM ranges from -1 to 1, where a value of 1 indicates perfect similarity, and 0 indicates no similarity. The SSIM is reported as $\text{SSIM}(\bar{A},A_t)$, with $\bar{A}$ the reference atlas to which $A_t$ is compared.

We reported the mean and standard deviation over the gestational days $t$ for $\text{HV}_{\text{error}}(A_t,t)$, $\text{sharpness}(A_t)$ and $\text{SSIM}(\bar{A},A_t)$.

\newpage
\subsection{Experiment 1: Ablation and hyperparameters studies}
To evaluate our contributions to take the rapid anatomical development of the embryonic brain into account in the atlas generation, we performed ablation and hyperparameter studies. To this end, we evaluated all combinations of $\lambda_{\text{atlas}} \in \{0,1,100\}$, where $\lambda_{\text{atlas}}=0$ means no penalization of generated deviations, and $\delta \in \{0,\infty\}$, where $\delta=\infty$ gives a time-independent constant initial atlas. Subsequently, we evaluated the effect of varying $\delta \in \{1,2,3,4\}$ compared to $\delta=0$, which increases the number of images used to create the initial atlas at every time point, leading to a smoother initial atlas. We used the default value $\lambda_{\text{constraint}}=10$, $\lambda_{\text{deformation}}=0.01$ since these gave the best results in earlier work \citep{Bastiaansen2022wbir}.  The average number of images per gestational day in the atlas-generation set was 18, hence we used $\kappa=18$. For all studies, we compared the results on the validation set. The model trained using the best value for $\delta$ on the validation set was applied to the test set.

\subsection{Experiment 2: Analysis best model}
The best model in Experiment 1 was evaluated as follows. Firstly, to assess how well the constraint for groupwise registration $\sum_{i,t} \phi_{i,t} \approx 0$ was satisfied, we calculated for every voxel $x \in S_{A_{t}}$ the average deformation $U_t(x)$ at every gestational day $t$:

\begin{equation} U_t(x) = \frac{1}{n(t)}\sum_{k=1}^{n(t)} u_{t}^k(x)  \end{equation}

with $n(t)$ the number of images at gestational day $t$, and $u_{t}^k$ the $k^\text{th}$ displacement field at day $t$. We analyzed the mean and standard deviation over all voxels $x \in S_{A_{t}}$ of the average deformation $U_t(x)$. Secondly, to assess the learned generated deviation $A_t^g$, we visually inspected the shape and position of ventricles and brain structures in $A_g^t$. Thirdly, we evaluated the influence of atlas-generation dataset size. Dataset size was evaluated by splitting the atlas-generation set into $n=2,4,8$ subsets. When $n=2$ the subsets consist of approximately 100 subjects with 300 images, when $n=4$ of 75 subjects with 150 images, and when $n=8$ of 35 subjects with 80 images. We call the atlases $A_t^{n,m}$, with $n$ the number of subsets and $m=1,...,n$ the subset index. The  $\text{SSIM}(A_t,A_t^{n,m})$ and  $\text{HV}_{\text{error}}(A_t^{n,m},t)$ for every combination of $n \in \{2,4,8\}$ and $m=1,...,n$ was calculated. Here,  $A_t$ is the atlas obtained using the complete atlas-generation set.

\subsection{Experiment 3: Visual comparison to the 3D Embryo Atlas and Fetal Brain atlas}
Since no ground truth exists for atlases, we choose to assess the presence of brain structures in the 4D Human Embryonic Brain Atlas by comparing it to the most similar atlases available. The first atlas used for comparison was the 3D Embryo Atlas developed by \citet{Bakker2016}, which covers the first trimester and consists of interactive 3D digital reconstructions based on histologically sectioned human embryos of the Carnegie collection.  The 3D Embryo Atlas contains Carnegie stage 18 (estimated GA in days: 58 - 62), 20 (estimated GA in days: 65 - 67), 21 (estimated GA in days: 67 - 69), and 23 (estimated GA in days: 70 - 74). For comparison, in the 4D Human Embryonic Brain Atlas, we selected the central day of the estimated GA window of each Carnegie stage.

The second atlas used for comparison was the Fetal Brain Atlas by \citet{Namburete2023}. The Fetal Brain Atlas consists of a set of templates covering gestational weeks 14 to 30 and is based on ultrasound imaging. Since the gestational age ranges of the Fetal Brain Atlas and our 4D Human Embryonic Brain Atlas do not overlap, we compared subject-specific ultrasound images to the atlases and visually inspected whether the same brain structures were visible in the subject-specific ultrasound images and atlases. Specifically, we compared a subject from the atlas-generation set imaged at gestational day 87 (12 weeks + 3 days) to the corresponding representation in the 4D Human Embryonic Brain Atlas at day 87. Similarly, a subject imaged at gestational day 101 (14 weeks + 3 days), taken from the Rotterdam Periconceptional cohort \citep{Rousian2021}, was compared to the Fetal Brain atlas at gestational week 14. 

The atlases were visually compared in terms of: 1) ventricle shape and 2) the presence of the choroid plexus and the cerebellum. To aid visual inspection, we manually segmented the ventricles (lateral ventricles, third ventricle, cavity of the mesencephalon, and fourth ventricle), choroid plexus, and cerebellum in the 4D Human Embryonic Brain Atlas.

\newpage
\subsection{Experiment 4: Influence of maternal BMI on morphology}
To assess whether the atlas can capture known differences in morphology, we studied
the influence of maternal overweight and obesity during early pregnancy on first trimester brain development using VBM. We compared the following two populations: 1) subjects with a normal maternal BMI ($19 \leq$ BMI $<25)$ and 2) subjects with a high maternal BMI (BMI $\geq 25$, overweight and obesity). Since there were only 7 subjects with a low BMI (BMI<19), we excluded those from this analysis. Additionally, for subjects who participated multiple times in the study, we retained only the first pregnancy to prevent including the same maternal environment twice. We tested whether there was a structural volumetric difference between the two populations by performing a t-test. Because maternal overweight and obesity may differently influence brain development, we additionally performed subgroup analyses, comparing normal BMI versus overweight ($25 \leq$ BMI $<30$), normal BMI versus obese (BMI $\geq 30$), and overweight versus obese.

Since the statistical test is performed in every voxel and in multiple time windows, we corrected for multiple testing by false discovery rate (FDR) controlling of 5\% \citep{Benjamini1995}. This resulted in a voxel-wise map per time window, highlighting the voxels with a structural volumetric difference between the two populations.

\section{Results}

\subsection{Experiment 1: Ablation and hyperparameters studies}
Table \ref{tab:exp1} shows the quantitative results of the ablation and hyperparameter studies on the validation set. Firstly, focusing on registration accuracy for all tested values of $\lambda_{\text{atlas}}$ and $\delta$, the DSC was around between 0.75 - 0.88, indicating good registration accuracy of the head. Additionally, approximately $2\%$ percentage of voxels had a non-positive Jacobian determinant. Upon visual inspection, folding appeared in the majority of the cases on the edge of the brain.

\begin{figure}[t!]
  \centering
  \includegraphics[width=\textwidth]{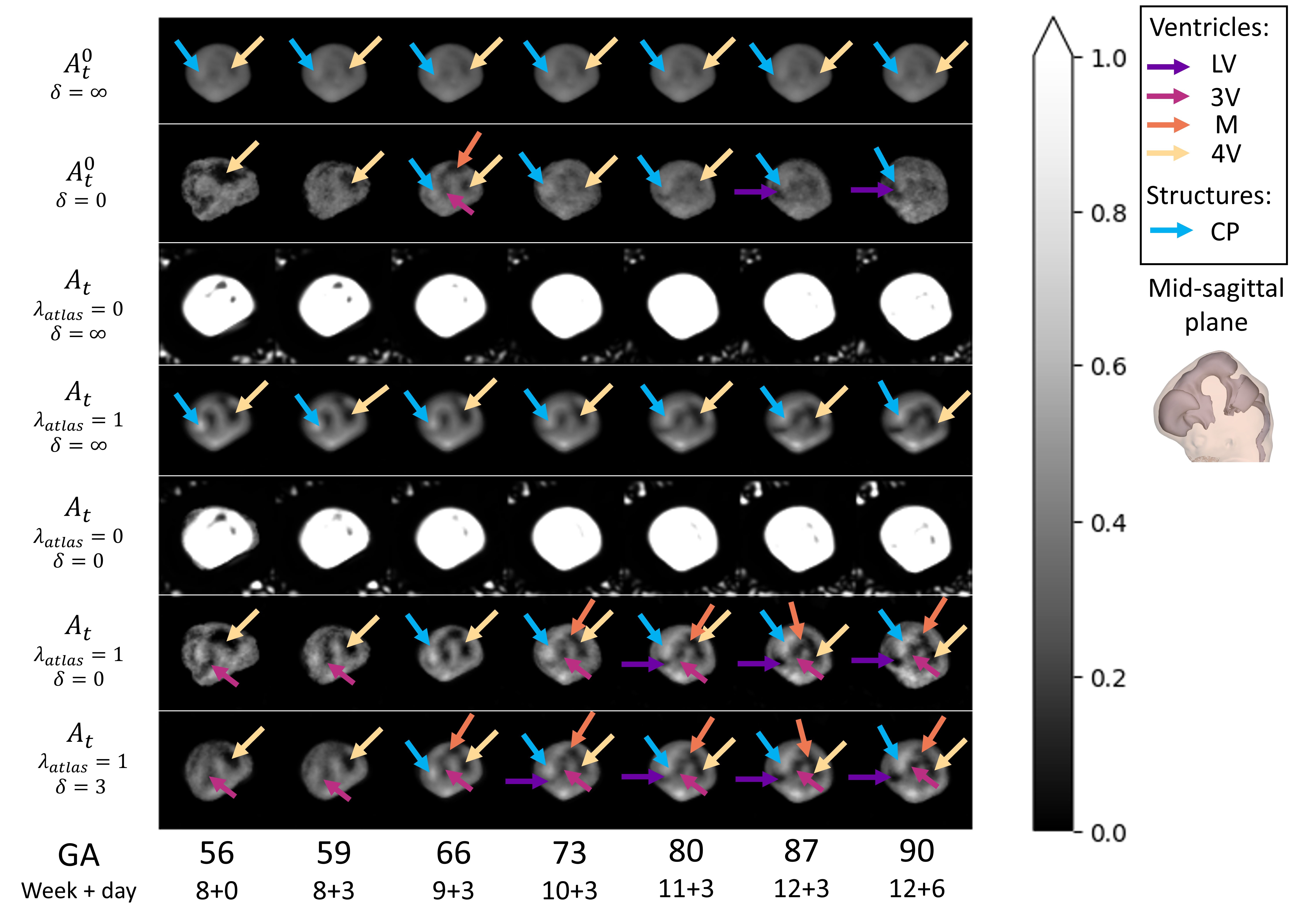}
  \caption{Atlas $A_t$ at different gestational days for different hyperparameter values tested in Experiment 1. All images are shown in the mid-sagittal plane. The arrows indicate relevant visible brain structures. LV = lateral ventricle, 3V = third ventricle (cavity of the diencephalon), M = cavity of the mesencephalon, 4V = fourth ventricle (cavity of the rhombencephalon), CP = choroid plexus, CER = cerebellum.}
  \label{fig:fig_exp1_exp2}
\end{figure}

Secondly, focusing on the ablation study for using a time-dependent initial atlas, we observed for $\delta=\infty$ (constant time-independent initial atlas) in Figure \ref{fig:fig_exp1_exp2} that the atlas contains anatomical inaccuracies. Notably, on gestational days 56 and 59, the fourth ventricle appears too small, and there seems to be an erroneous depiction of the choroid plexus. Similar differences in anatomy can be observed in the initial atlases for $\delta=\infty$ and $\delta=0$ in Figure \ref{fig:fig_exp1_exp2}. 

Thirdly, focusing on the ablation study for constraining deviations from the initial atlas (setting $\lambda_{\text{atlas}}>0)$, we observed that $\lambda_{\text{atlas}}=0$ gives a generated atlas showing incorrect anatomy, regardless of the value of $\delta$. Additionally, the SSIM between the initial atlas $A_t^0$ and the generated atlas $A_t$ was 0.174 and 0.206, indicating low similarity. For $\lambda_{\text{atlas}}=100$, the SSIM almost equaled 1 for both values of $\delta$, indicating that $A_t^0$ and $A_t$ were very similar. If $A_t^0$ and $A_t$ were very similar, the atlas generator network could not improve upon $A_t^0$. When choosing $\lambda_{\text{atlas}} = 1$, we found an SSIM of 0.876 and 0.867, and we observed the ventricles more clearly in $A_t$ compared to $A_t^0$, as shown in Figure \ref{fig:fig_exp1_exp2}. Hence, from the ablation study, we conclude that both the generated deviation from the initial atlas should be constrained, using $\lambda_{\text{atlas}}=1$, and a time-dependent initial atlas, using $\delta=0$, are needed to generate an anatomical correct atlas.

\begin{figure}[t!]
  \centering
  \includegraphics[width=0.5\textwidth]{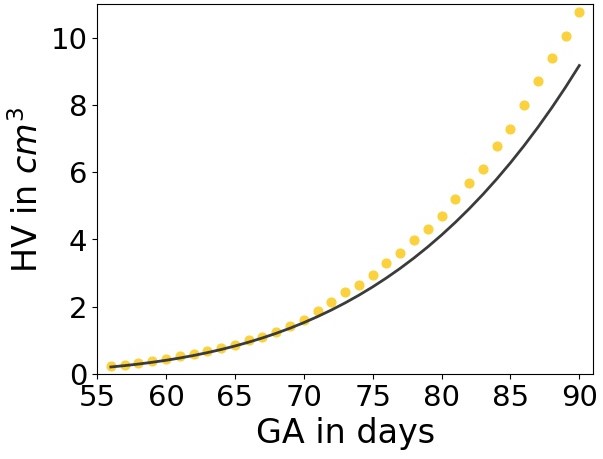}
  \caption{HV-curve for $\delta=3$ in Experiment 1. The gray line is the reference HV-curve $\text{HV}_{\text{VR}}(t)$ derived in previous research \citep{Koning2016} and the dots represent $V(A_t)$ the HV for every gestational day of the atlas $A_t$.}
  \label{fig:fig_hvcurve}
\end{figure}

Fourthly, we evaluated the hyperparameter $\delta \in {1,2,3,4}$. Increasing $\delta$ reduced the $\text{HV}_{\text{error}}(A_t,t)$, while no clear trend was observed for $\text{sharpness}(A_t)$. However, sharpness was slightly higher for $\delta=3$ than for $\delta=4$. Visual inspection revealed that for $\delta=4$, the choroid plexus appeared as early as 56 days, similar to $\delta=\infty$, which is anatomically incorrect (see Supplementary material 2 Figure~\ref{fig:figure_supp_2}). We therefore selected $\delta=3$, as it provided a good balance between volumetric accuracy, image sharpness, and anatomical plausibility. As shown in Figure~\ref{fig:fig_exp1_exp2}, the atlas at $\delta=3$ appears less noisy, and Figure~\ref{fig:fig_hvcurve} demonstrates that $\text{V}(A_t)$ increases monotonically, indicating temporally consistent volumetric growth.

Finally, in Table \ref{tab:exp1} the resulting evaluation metrics for the best hyperparameter over the test set are given, comparable to the results found on the validation set.

\begin{table}[b!]
 \caption{Results for the comparison of hyperparameters in Experiment 1. For $\% | \nabla J | \leq 0$ and $\text{DSC}$ the mean and standard deviation over the validation and/or test set are given. For $\text{SSIM}(A_t^0,A_t)$, $\text{HV}_{\text{error}}(A_t,t)$ and $ \text{sharpness}(A_t)$ the mean and standard deviation over all gestational days $t$ is given.}
  \centering
  \begin{tabular}{ccccccc}
     \midrule
    $\mathbf{\lambda_{\text{atlas}}}$ & $\mathbf{\delta}$ & $\mathbf{\% | \nabla J | \leq 0} \downarrow $ & \textbf{DSC} $\uparrow$ &$\mathbf{\text{\textbf{SSIM}}(A_t^0,A_t)} \uparrow$ & $\mathbf{\text{\textbf{HV}}_{\text{\textbf{error}}}(A_t,t)}(\%) \downarrow$ &$ \mathbf{\text{\textbf{sharpness}}(A_t)} \uparrow$ \\
    \midrule
    \multicolumn{2}{l}{\textbf{Validation set}} &&& \\
    0 & $\infty$ &2.40 (0.83) & 0.755 (0.173)& 0.174 (0.014) & 14.4 (4.69) & 0.397 (0.012) \\
    1 & $\infty$ & 3.56 (1.31) & 0.860 (0.155) & 0.876 (0.024) & 13.9 (4.7) & 0.241 (0.018) \\
    100 & $\infty$ & 3.43 (1.31) & 0.875 (0.153) & 0.992 (0.002)  & 8.6 (5.0)  & 0.408 (0.011) \\
    0 & 0 & 2.56 (1.00) & 0.862 (0.155)  & 0.206 (0.019) & 27.6 (4.4) & 0.362 (0.009)\\
    1 & 0 & 1.91 (0.81) & 0.825 (0.151)  & 0.867 (0.023) & 15.4 (7.8)& 0.307 (0.073)\\
    100 & 0 & 1.51 (0.87) & 0.832 (0.149)  & 0.989 (0.003)& 13.5 (8.5)& 0.493 (0.021)\\
    \midrule
    \multicolumn{2}{l}{\textbf{Validation set}} &&& \\
    1 & 1 &  1.81 (0.97) & 0.878 (0.148) & 0.908 (0.011)  & 13.1 (5.3)& 0.338 (0.012)\\
    1 & 2 & 2.14 (1.08) & 0.880 (0.151) & 0.913 (0.008) & 12.4 (5.1) & 0.278 (0.023)\\
    1 & 3 & 2.41 (1.10) & 0.877 (0.149)  & 0.916 (0.009) &  12.2 (4.5)& 0.306 (0.020)\\
    1 & 4 & 2.82 (1.23) & 0.878 (0.152) & 0.916 (0.152) & 11.5 (4.2)  & 0.286 (0.017)\\
    \midrule
    \multicolumn{2}{l}{\textbf{Test set}} &&& \\
    1 & 3 & 2.36 (1.10) & 0.886 (0.135) & - & - &- \\
     \midrule
  \end{tabular}
  \label{tab:exp1}
\end{table}

\clearpage

\subsection{Experiment 2: Analysis best model}
\begin{figure}[b!]
  \centering
  \includegraphics[width=0.9\textwidth]{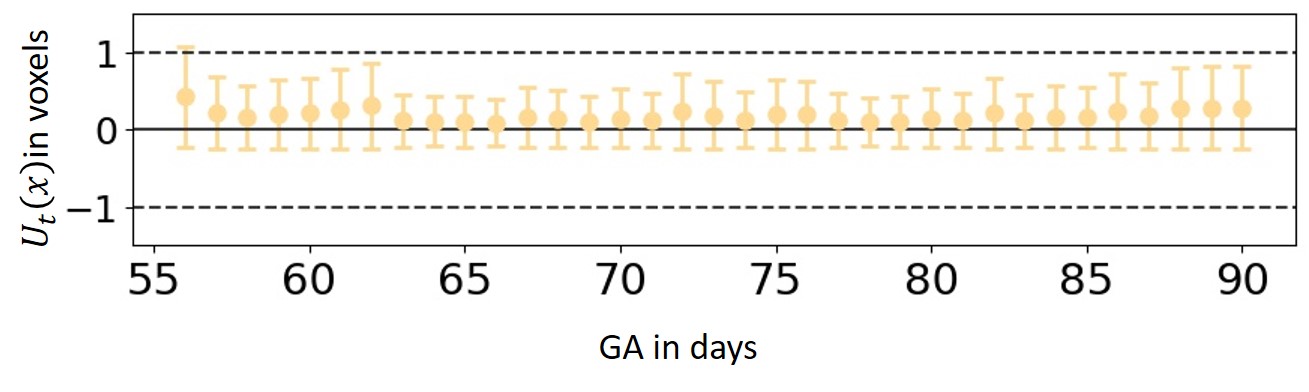}
  \caption{Assessment of the groupwise constraint in Experiment 2. The distribution of the average displacement $U_t(x)$ over all voxels $x\in S_{A_{t}}$ per gestational day $t$, given in voxels. The error bar indicates the standard deviation over all voxels, the solid line indicates perfect fulfillment of the constraint, and the dashed line indicates the bounds for sub-voxel displacement.}
  \label{fig:fig_exp2_groupwise}
\end{figure}

First, we assessed how well the groupwise constraint was satisfied. The results are shown in Figure \ref{fig:fig_exp2_groupwise}, where we observed that the average displacement is close to zero and all error bars, except at gestational day 56, fall within the bounds for sub-voxel average displacement. For days with more images available (around days 65 and 80 in weeks 9 and 11) the standard deviation was smaller. Around the edges at days 56 and 90 the largest spread was visible.

\begin{figure}[b!]
  \centering
  \includegraphics[width=\textwidth]{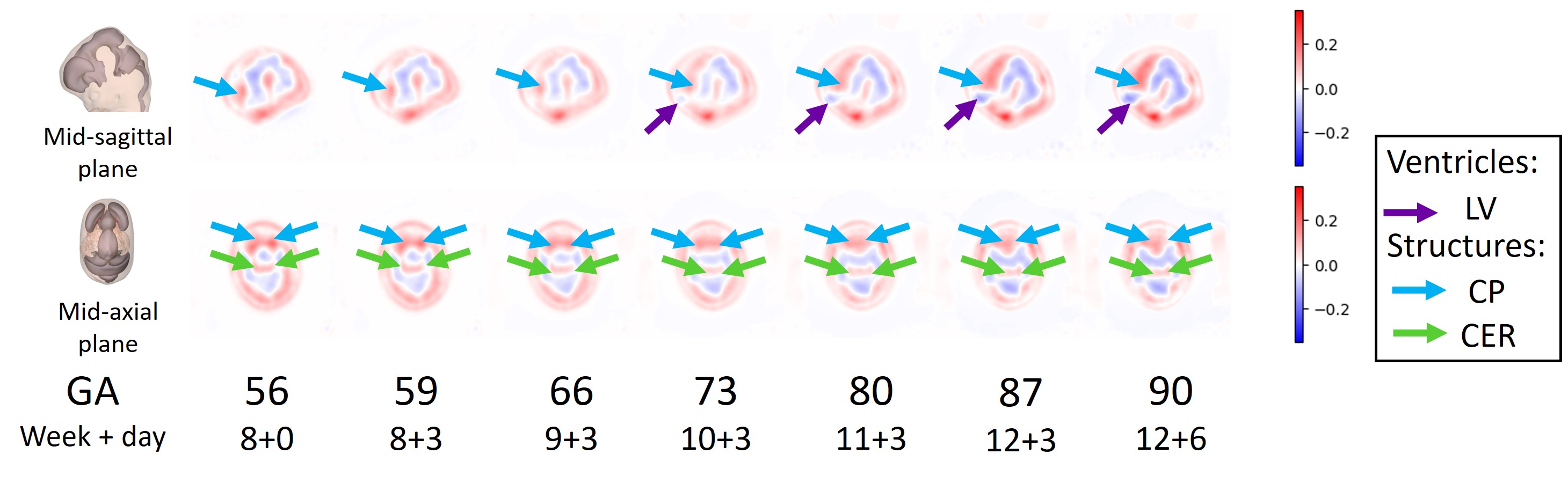}
  \caption{Generated deviation $A_t^g$ at different gestational days in Experiment 2. The arrows indicate deviations within relevant visible brain structures. LV = lateral ventricle, CP = choroid plexus, CER = cerebellum.}
  \label{fig:fig_exp2_gen_atlas}
\end{figure}

In Figure \ref{fig:fig_exp2_gen_atlas}, the generated deviation $A_t^g$ for different gestational days is given in the mid-sagittal and mid-axial plane. We observed that the brain ventricles became more hypo-intense and the brain structures such as the choroid plexus and the cerebellum more hyper-intense compared to $A_t^0$. In the mid-sagittal plane, the growth of the choroid plexus is clearly visible, which starts in the anterior of the brain and grows posterior. In $A_t^g$ in mid-axial plane,  the morphological changes of the cerebellum and the appearance of the choroid plexus can be observed clearly.  Furthermore, in the $A_t^g$ the lateral ventricles are visible around the choroid plexus, which is not always the case in the images $I_{i,t}$ due to the small size of the lateral ventricles and the noise present in ultrasound imaging.

\begin{figure}[b!]
  \centering
  \includegraphics[width=0.9\textwidth]{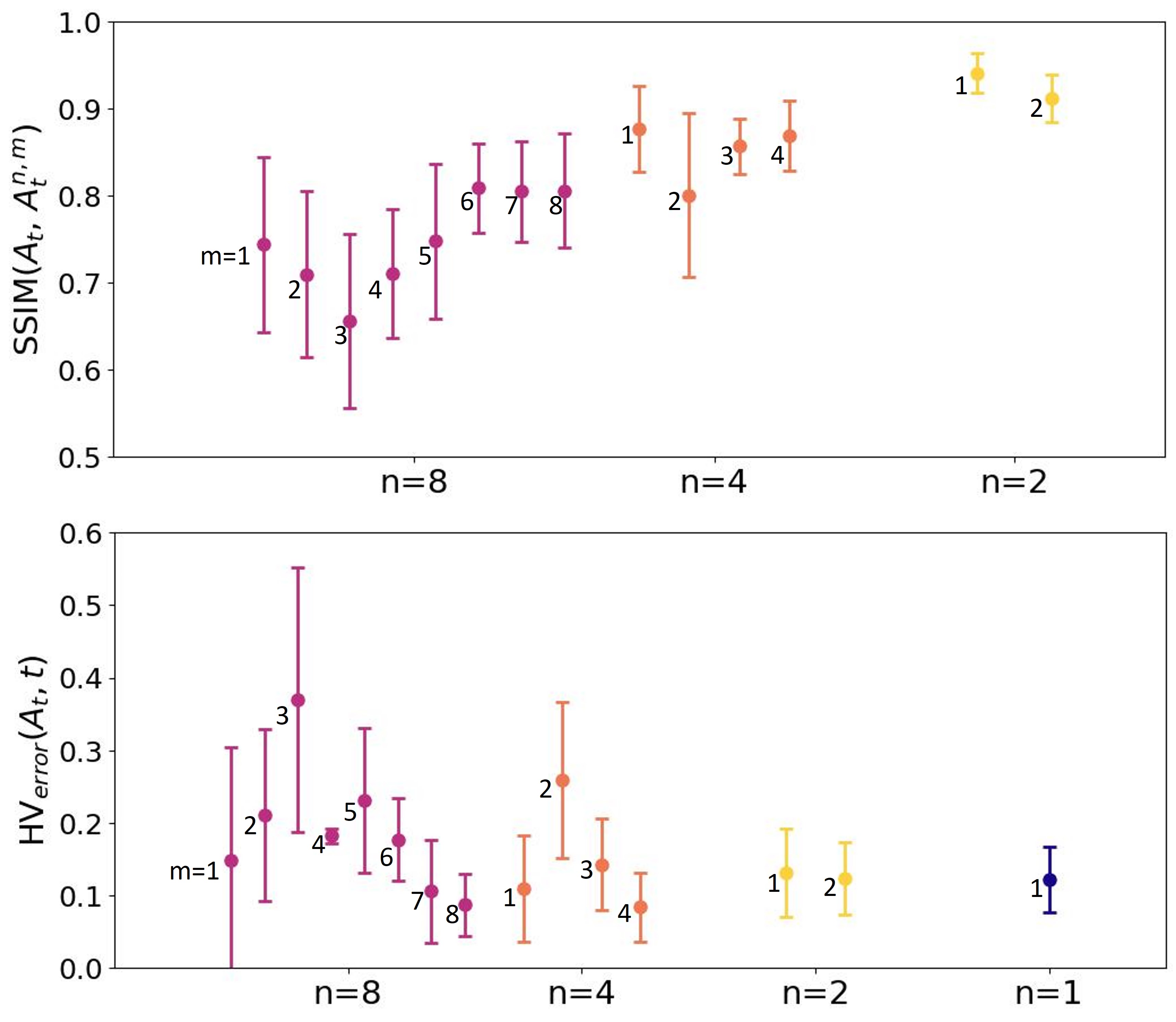}
  \caption{Evaluation of the influence of atlas-generation dataset size in Experiment 2. Shown are the $\text{SSIM}(A_t, A_t^{n,m})$ and $\text{HV}_{\text{error}}(A_t^{n,m},t)$ with the number of subsets $n \in \{2,4,8\}$ and subset index $m=1,...,n$. Dots indicate the mean and error bars indicate the standard deviation.}
  \label{fig:fig_exp3}
\end{figure}

\begin{figure}[t!]
  \centering
  \includegraphics[width=\textwidth]{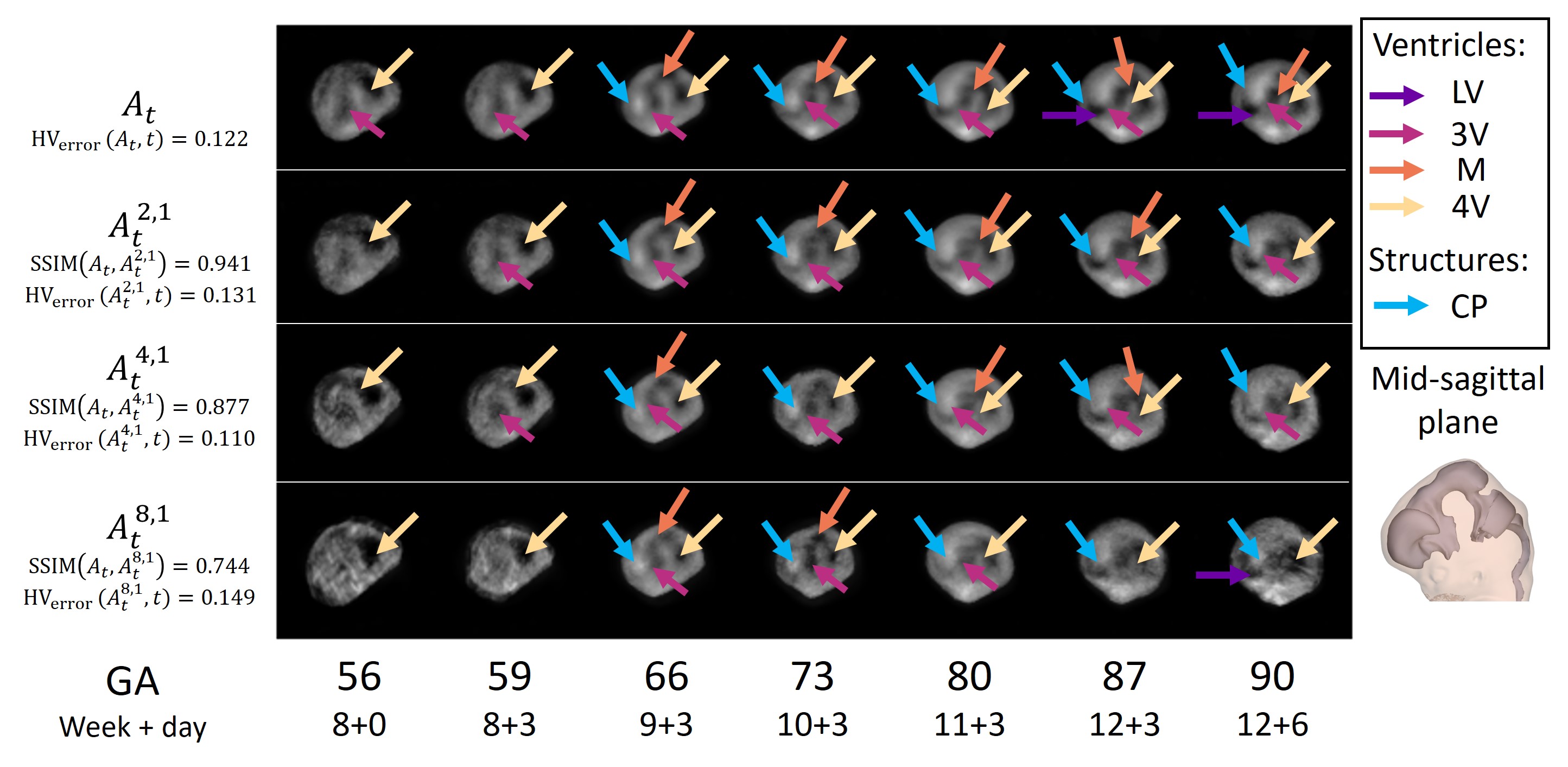}
  \caption{$A_t$ at different gestational days for different atlas-generation set sizes from Experiment 2, shown in the mid-sagittal plane. The arrows indicate relevant visible brain structures. LV = lateral ventricle, 3V = third ventricle (cavity of the diencephalon), M = cavity of the mesencephalon, 4V = fourth ventricle (cavity of the rhombencephalon), CP = choroid plexus, CER = cerebellum.}
  \label{fig:fig_exp3visual}
\end{figure}

Figure \ref{fig:fig_exp3} gives the $\text{SSIM}(A_t,A_t^{n,m})$ and $\text{HV}_{\text{error}}(A_t^{n,m},t)$ for $n \in \{2,4,8\}$ and $m=1,...,n$. The error bars indicate the standard deviation over the gestational days for each atlas $A^{n,m}_t$. For both metrics, we observed that when $n$ gets bigger the metric is impacted negatively, and the spread over time gets wider. Furthermore, the variation among the $m$ generated atlases increased as $n$ increased. Figure \ref{fig:fig_exp3visual} shows a visualization of the different atlases. Despite the lower SSIM as $n$ increased, in most cases, structures remained visible within the brain. The decrease in SSIM was mainly due to noise.

\clearpage
\subsection{Experiment 3: Visual comparison to the 3D Embryo Atlas and Fetal Brain atlas}
\begin{figure}[b!]
  \centering
  \includegraphics[width=\textwidth]{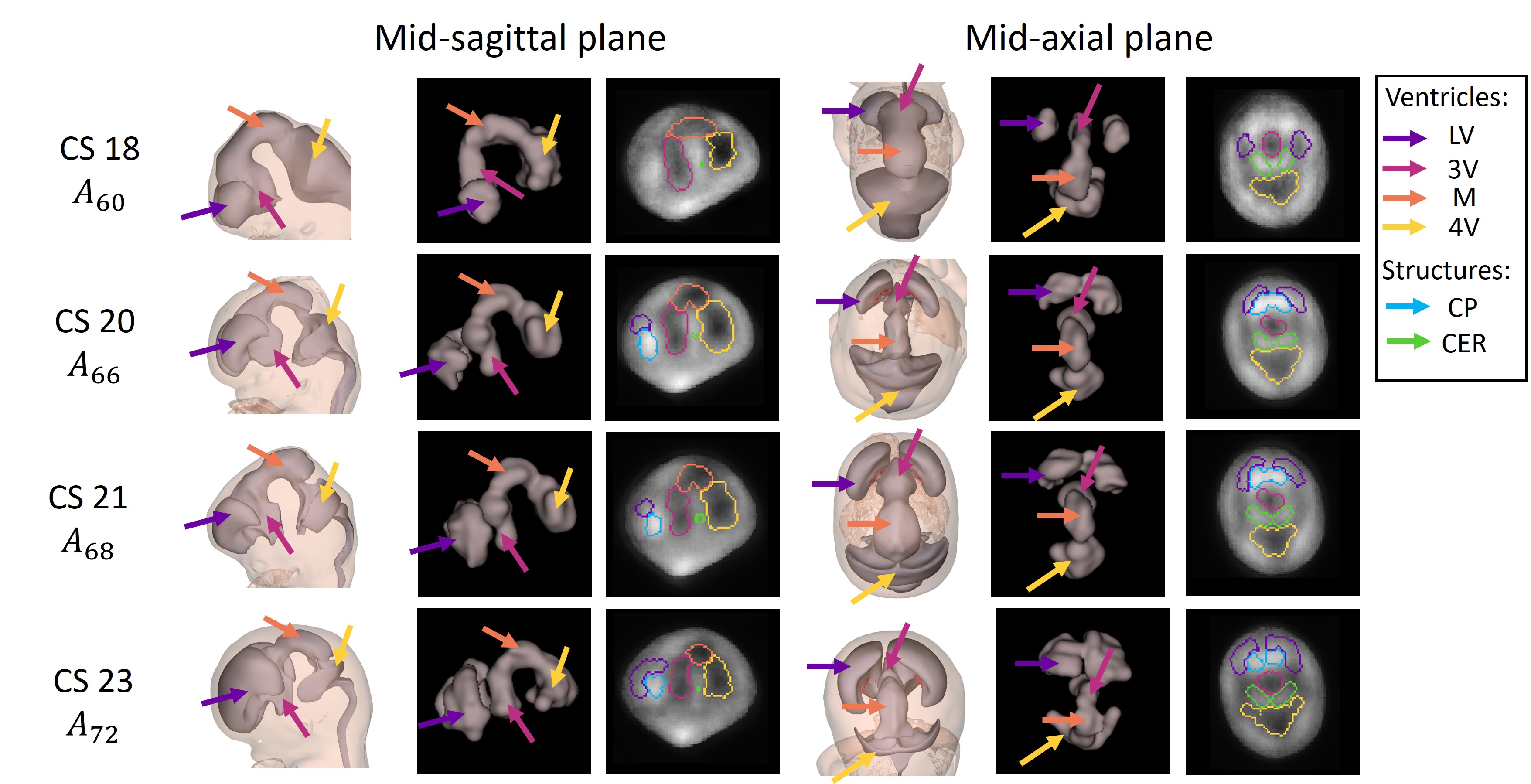}
  \caption{Comparison of the 4D Human Embryonic Brain Atlas $A_t$ and the 3D Embryo Atlas in Experiment 3. The 3D renderings are given in the mid-sagittal and mid-axial plane. The 4D Human Embryonic Brain Atlas $A_t$ is also shown in mid-sagittal and mid-axial plane. The colors of the delineation match the colors of the arrows. CS = Carnegie stage, LV = lateral ventricle, 3V = third ventricle (cavity of the diencephalon), M = cavity of the mesencephalon, 4V = fourth ventricle (cavity of the rhombencephalon), CP = choroid plexus, CER = cerebellum.}
  \label{fig:fig_exp4_amc}
\end{figure}

Firstly, we compared the 4D Human Embryonic Brain Atlas to the 3D Embryo Atlas \citep{Bakker2016}. Figure \ref{fig:fig_exp4_amc} shows the mid-sagittal and mid-axial plane of both atlases.In both atlases, the 3D renderings of the ventricles have similar shapes at each GA, with clearly visible barriers between the different regions of the ventricles. Furthermore, in the 3D Embryo Atlas at Carnegie stage 20 the choroid plexus becomes visible, which can also be clearly delineated in the 4D Human Embryonic Brain Atlas. 

\begin{figure}[t!]
  \centering
  \includegraphics[width=\textwidth]{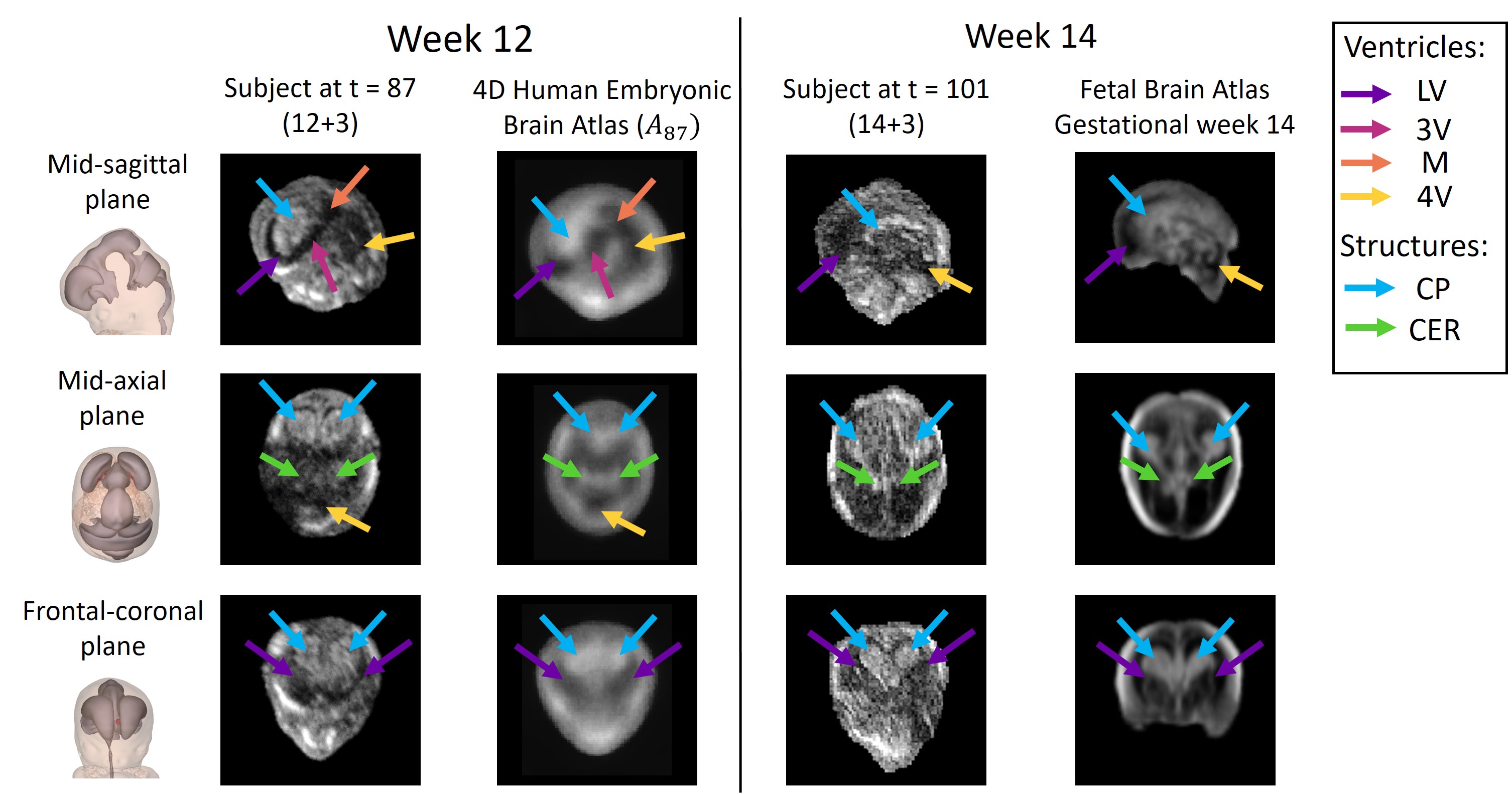}
  \caption{Comparison of a subject at gestational day 87 and the 4D Human Embryonic Brain Atlas $A_{87}$, and of a subject at gestational day 101 and the Fetal Brain Atlas in gestational week 14 in Experiment 3. The mid-sagittal, mid-axial, and frontal-coronal plane. CS = Carnegie stage, LV = lateral ventricle, 3V = third ventricle (cavity of the diencephalon), M = cavity of the mesencephalon, 4V = fourth ventricle (cavity of the rhombencephalon), CP = choroid plexus, CER = cerebellum.}
  \label{fig:fig_exp4_oxford}
\end{figure}

Secondly, we compared a subject imaged at gestational day 87 to the 4D Human Embryonic Brain Atlas at day 87, and a subject imaged at gestational day 101 to the Fetal Brain Atlas in gestational week 14 \citep{Namburete2023}. Both comparisons in Figure \ref{fig:fig_exp4_oxford} show in both subject-specific ultrasound images and the atlases similarly shaped ventricles, the choroid plexus, and the cerebellum. Differences in morphology due to aging are clearly visible when comparing the two atlases, for example, the more posterior position of the choroid plexus in gestational week 14. These differences are in line with the differences observed when comparing the subject-specific ultrasound images. 

Comparing the 4D Human Embryonic Brain Atlas to the Fetal Brain Atlas, the Fetal Brain Atlas shows more details and clearer structures. Another clear difference between the atlases is how the brain was extracted. In the Fetal Brain Atlas, the atlas by \citet{Gholipour2007} was used to extract the brain. Since no (MRI) atlas covering the first trimester was present (see Table \ref{tab:relatedwork}), we extracted the brain by segmenting the embryonic head, which resulted in an atlas that includes the early facial structures.

\clearpage

\subsection{Experiment 4: Influence of maternal BMI on morphology}
\begin{figure}[t!]
  \centering
  \includegraphics[width=\textwidth]{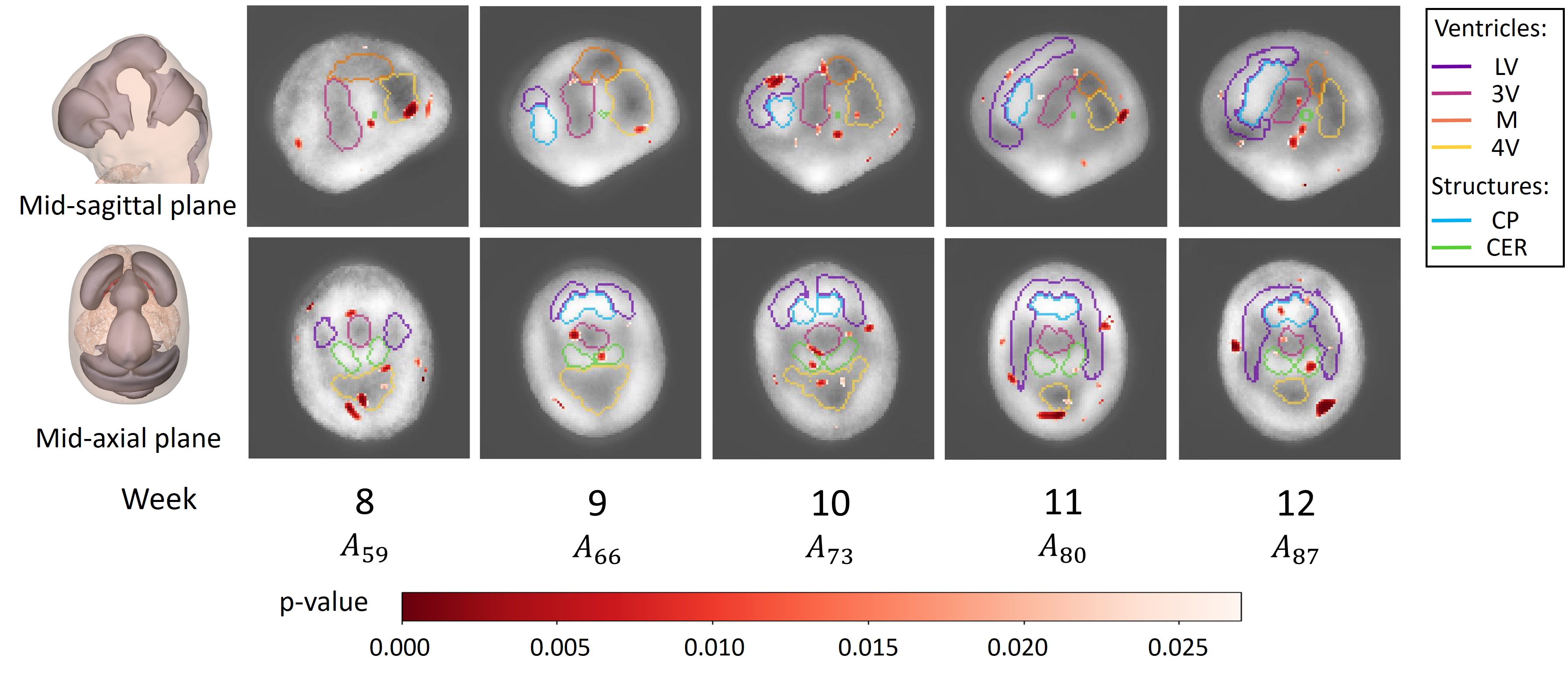}
  \caption{Visualization of the comparison between brain morphology for the normal (control) and high BMI group for every week GA using VBM in Experiment 4. The red voxels indicate the p-value for significant structural volumetric differences (p<0.027). The atlas of the middle of every gestational week is shown. LV = lateral ventricle, 3V = third ventricle (cavity of the diencephalon), M = cavity of the mesencephalon, 4V = fourth ventricle (cavity of the rhombencephalon), CP = choroid plexus, CER = cerebellum.}
  \label{fig:fig_exp5_visual}
\end{figure}

Applying the $5\%$ FDR correction, we found that $p<0.027$ was significant in all the analyzes. Using $\delta=3$ from Experiment 1, the GA window became 7 days. Figure \ref{fig:fig_exp5_visual} shows significant voxels in the mid-sagittal and mid-axial plane for each gestational week after performing VBM in red when comparing normal maternal BMI to high maternal BMI. The atlas and delineated regions correspond to the middle day of each week (days 59, 66, 73, 80, 87). Note that most significant differences occurred within or near the delineated regions. Starting at week 9, these significant differences were present in the cerebellum in the mid-axial plane, which was expected since in previous work a negative association was found between maternal BMI and trans-cerebellar diameter \citep{Koning2016}. There were also significant differences in non-delineated regions. For example in weeks 8, 11, and 12 in the axial plane, there was a hyper-intense region with significant p-values close to the fourth ventricle near the developing brain stem.

\begin{figure}[b!]
  \centering
  \includegraphics[width=0.5\textwidth]{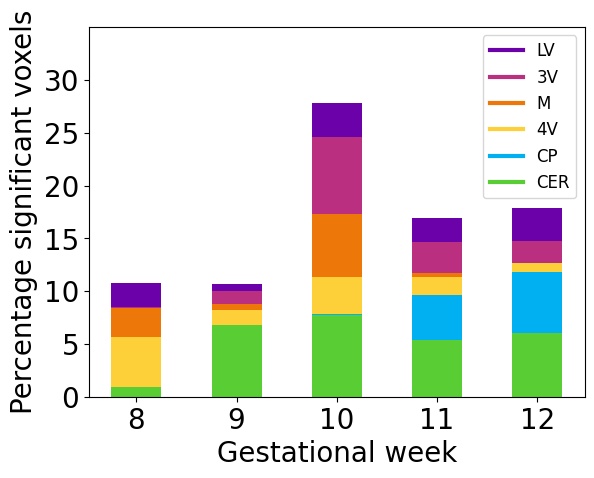}
  \caption{Percentage of significantly different voxels within the delineated ventricles and brain structures at different gestational weeks for the VBM analysis in Experiment 4. LV = lateral ventricle, 3V = third ventricle (cavity of the diencephalon), M = cavity of the mesencephalon, 4V = fourth ventricle (cavity of the rhombencephalon), CP = choroid plexus, CER = cerebellum.}
  \label{fig:fig_exp5}
\end{figure}

Figure \ref{fig:fig_exp5} shows the percentage of voxels with a significant p-value within every delineated structure per gestational week. Considering both figures, we observed that the region where the most significant differences appear is the cerebellum, followed by the fourth ventricle. Starting in gestational week 11, significant differences appear in the developing choroid plexus. Significant differences are present in the lateral ventricles throughout the first trimester, the cavity of the mesencephalon shows more significant differences prior to gestational week 11, and the third ventricle show more significant differences starting at gestational week 10.

We additionally performed subgroup analyses comparing normal BMI vs. overweight, normal BMI versus obese, and overweight versus obese. The normal versus overweight results largely overlapped with the combined high-BMI analysis and also showed additional differences near the fourth ventricle at 12 weeks (Supplementary Figures \ref{fig:figure_supp_3}A and B and \ref{fig:figure_supp_4}A and B). The normal versus obese comparison revealed differences in the mesencephalon that were less prominent when overweight and obesity were combined (Supplementary Figures \ref{fig:figure_supp_3}A and C and \ref{fig:figure_supp_4}A and C). The overweight versus obese analysis most clearly highlighted subgroup-specific effects, showing differences in the mesencephalon at week 10 and the fourth ventricle at week 12, consistent with the separate analyses (Supplementary Figures \ref{fig:figure_supp_3}D and \ref{fig:figure_supp_4}D).

\clearpage
\section{Discussion}
In the first trimester, the majority of congenital brain anomalies are missed during ultrasound examination, due to the sonographers' lack of detailed knowledge of the physiological appearance of the rapidly developing and changing embryonic brain \citep{Volpe2021, Buijtendijk2024}. Less severe but still relevant are the early morphological differences caused by maternal conditions such as age and obesity, and lifestyle behavior, such as nutrition, smoking, and alcohol use \citep{Koning2016,Koning2017}. However, these differences are even harder or impossible to visually assess. To offer a comprehensive view of development, we developed the 4D Human Embryonic Brain Atlas. This spatiotemporal brain atlas was constructed using 654 ultrasound images of 402 subjects obtained between 8 and 12 weeks of gestational age, employing a deep learning approach for groupwise image registration and atlas creation. 

Our method introduces a time-dependent initial atlas and penalizes deviations to preserve age-specific anatomy during rapid first trimester development. While building on \citet{Dalca2019}, which was evaluated in the context of Alzheimer’s disease and characterized by relatively subtle anatomical changes, our approach explicitly models the pronounced and rapid anatomical variations of early brain development. The ablation study demonstrated that both adaptations were essential for anatomical accuracy, with their omission resulting in incorrect atlas anatomy, highlighting the necessity of these contributions for generating age–specific atlases that maintain anatomical consistency over time.

We evaluated the resulting 4D Human Embryonic Brain Atlas in terms of volumetric growth and anatomy. We found that the HV-curve of the atlas had an average relative difference of $12.2 \%$ with respect to the curve obtained in previous research by \citet{Koning2016}. This difference may partially be explained by the study population used to create the HV-curve. Namely, IVF/ICSI pregnancies were included, which has been shown to influence volumetric growth \citep{Koning2016}. In terms of anatomy, the visual comparison between the 4D Human Embryonic Brain Atlas and the 3D Embryo Atlas \citep{Bakker2016} demonstrated a similar appearance of brain structures and ventricles. The comparison to the Fetal Brain Atlas \citep{Namburete2023} showed the same brain structures and ventricle shaped in subject-specific ultrasound images and the atlases, indicating that the 4D Human Embryonic Brain Atlas reflects the structural information present in the subject-specific ultrasound images similar to the Fetal Brain Atlas. Finally, using VBM, we assessed whether the atlas could capture known differences in morphology. To this end, we studied the influence of maternal overweight and obesity during early pregnancy on first trimester brain development. Using VBM, we found that the cerebellum significantly differs in morphology in case of a high maternal BMI, which is in line with previous research \citep{Koning2017}.

Several limitations should be considered. In the VBM study, statistical tests were performed at every voxel (>2 million) within the brain. This may lead to false positives despite correction for multiple testing and make the results difficult to interpret, particularly in non-delineated regions. Furthermore, the analysis was performed on the entire embryonic head rather than the isolated brain, following the definition of the HV used in previous work \citep{Koning2016}. As non-brain tissue cannot yet be clearly separated during the first trimester and no standardized segmentation method exists, this ensured anatomical completeness but may have reduced interpretability in some regions. Future work should aim to segment individual brain regions to enable more localized analyses. In addition, the application of dimensionality reduction techniques, such as PCA \citep{Kurita2019}, could reduce the risk of false positives and facilitate interpretation by extracting the most relevant morphological variations among subjects.

A further consideration is the grouping of overweight and obese participants in the primary analysis. Our subgroup analyses showed that overweight and obesity separately led to partly distinct spatial patterns, some of which were less apparent when the categories are combined. These differences should be interpreted with caution given the limited sample size, particularly in the obesity group, and future studies are needed to validate and further explore these differences.

Another limitation is that we did not model the longitudinal relationship between images acquired at multiple time points during the first trimester. Although several participants underwent repeated scans, these were treated as independent samples. Future work could incorporate longitudinal information directly into the atlas generation process to better capture individual growth trajectories. Existing atlas methods listed in \ref{tab:relatedwork} that include longitudinal data still treat these images as cross-sectional, underscoring the challenges of capturing individual growth trajectories. Approaches that account for intra-subject trajectories often use a two-stage process: first constructing subject-specific atlases by registering intra-subject data, then creating a population-level atlas. This typically requires carefully designed datasets with imaging data available for all subjects at each time point \citep{Liao2012,Li2015}. A promising direction for future work is to combine the proposed time-dependent initial atlas with an implicit neural representation framework, enabling direct modeling of subject-specific developmental trajectories \citep{Chen2025}.

Compared with the Fetal Brain Atlas \citep{Namburete2023}, our atlas shows less detail and sharper structures. This might be explained by the several additional image processing steps taken, including speckle noise reduction and manually annotated landmarks, to improve the sharpness of brain structures in the Fetal Brain Atlas. In future work, besides including additional (manual) image processing steps, generative approaches such as latent diffusion models can be explored to improve atlas generation since they often produce sharper and more detailed images than registration-based approaches \citep{Rombach2022}. Another potential direction for future research to enhance visible details and brain structures is to refine the creation of the initial atlas. In the current approach, all images within the $\delta$-window are treated as equally important when generating the initial atlas. A possible extension could involve using (weighted) kernel regression, a technique commonly applied in non-learning-based methods \citep{Kuklisova2011, Serag2012, Gholipour2007}.

A limitation of the presented atlas is that it is based on ultrasound data acquired within the Rotterdam Periconceptional cohort \citep{Rousian2021,Steegers2016}. The data was acquired at fixed time points and following a standardized protocol. However, all patients included in the cohort were recruited from a tertiary university hospital, which may reduces the external validity of our results to the general population. Hence, validation within patients from the general population is needed.

Finally, regarding the clinical impact of the 4D Human Embryonic Brain Atlas, firstly, as demonstrated, the atlas can be used to analyze the influence of paternal and maternal conditions and lifestyle behaviors on brain development. Using a similar analysis, early detection of (congenital) anomalies and adverse growth patterns may be performed by comparing new subjects to the atlas. Secondly, the atlas holds significant value as a visualization aid for soon-to-be parents in clinical practice. It offers a unique and detailed perspective on the progression of early brain growth. Research has shown that confirmation of healthy development as shown during a standard ultrasound scan promotes parental attachment \citep{Walsh2014,Boukydis2009,Lapaire2007,Omalley2020}. Moreover, studies investigating novel visualization techniques of first trimester ultrasonography, such as offering an additional 3D VR visualization or using real-time ultrasound, observe a positive parental response and lead to reduced anxiety \citep{Pietersma2022,Lapaire2007,Popovici2019}.

\section{Conclusion}

We proposed a deep learning approach for spatiotemporal atlas generation capable of accurately capturing age-specific morphological features during rapid first trimester brain development. Using this approach, we generated the 4D Human Embryonic Brain Atlas, the first spatiotemporal atlas of the human embryonic brain covering days 56 to 90 of gestation. The atlas provides a framework for quantitative analysis of early brain development, enabling the study of parental and lifestyle factors, supporting early detection of congenital anomalies and abnormal growth patterns, and serving as a clinically valuable visualization aid for parents and healthcare providers. Overall, the 4D Human Embryonic Brain Atlas offers a detailed and accurate picture of early brain development, which may contribute to earlier detection, prevention, and treatment of neurodevelopmental disorders while facilitating further research in this critical developmental period.

\subsection*{Acknowledgements}
We thank all participants of the Rotterdam Periconceptional cohort (Predict study), without whom this study could not have been performed. Further, we would like to thank the Rotterdam Periconceptional Cohort team for their support in the processes of recruitment and data collection, which contributed to the results of this study. 

\subsection*{Ethics}
This study was approved by the local Medical Ethical and Institutional Review Board of the Erasmus MC, University Medical Center, Rotterdam, The Netherlands. Prior to participation, all participants provided written informed consent.

\subsection*{Data and code availability}
The code and 4D Human Embryonic Brain Atlas are publicly available at: \url{https://gitlab.com/radiology/prenatal-image-analysis/4D_human_embryonic_brain_atlas}.

\subsection*{Author Contributions}
\textbf{Wietske A.P. Bastiaansen:} Conceptualization, Methodology, Formal analysis, Data Curation, Writing - Original Draft 

\textbf{Melek Rousian:} Conceptualization, Formal analysis, Writing - Review \& Editing, Supervision, Funding acquisition

\textbf{Anton H.J. Koning:} Formal analysis, Writing - Review \& Editing

\textbf{Wiro J. Niessen:} Conceptualization, Writing - Review \& Editing, Supervision, Funding acquisition

\textbf{Bernadette S. de Bakker} Formal analysis, Writing - Review \& Editing

\textbf{Régine P.M. Steegers-Theunissen:} Conceptualization, Writing - Review \& Editing, Supervision, Funding acquisition

\textbf{Stefan Klein:} Conceptualization, Methodology, Formal analysis, Writing - Review \& Editing, Supervision, Funding acquisition

\subsection*{Declaration of Competing Interests}
Wiro J. Niessen is the founder, and was the scientific lead and stockholder of Quantib BV. Wiro J. Niessen was a board member of the Technical Branch of the Dutch Science Foundation (NWO-TTW) until January 2023. Wietske A.P. Bastiaansen, Melek Rousian, Régine Steegers-Theunissen, and Stefan Klein collaborate with the Women’s Health Ultrasound business unit of GE Healthcare, in two public-private collaboration projects (payment to institution). All other authors declare no conflicts of interest.

\subsection*{Declaration of generative AI and AI-assisted technologies in the writing process.}
During the preparation of this work, the authors used a large language model in order to improve readability and language. After using this tool, the authors reviewed and edited the content as needed and take full responsibility for the content of the publication.

\subsection*{Role of the funding source}
Wietske A.P. Bastiaansen is funded by Erasmus MC Medical Research Advisor Committee (grant number: FB 379283). The funders had no role in this study.

\bibliographystyle{plainnat}
\bibliography{references}

\appendix  
\newpage
\section{Supplementary material 1: Influence of $\delta$ during gestational week 8}
\begin{figure}[h!]
    \centering
    \includegraphics[width=\linewidth]{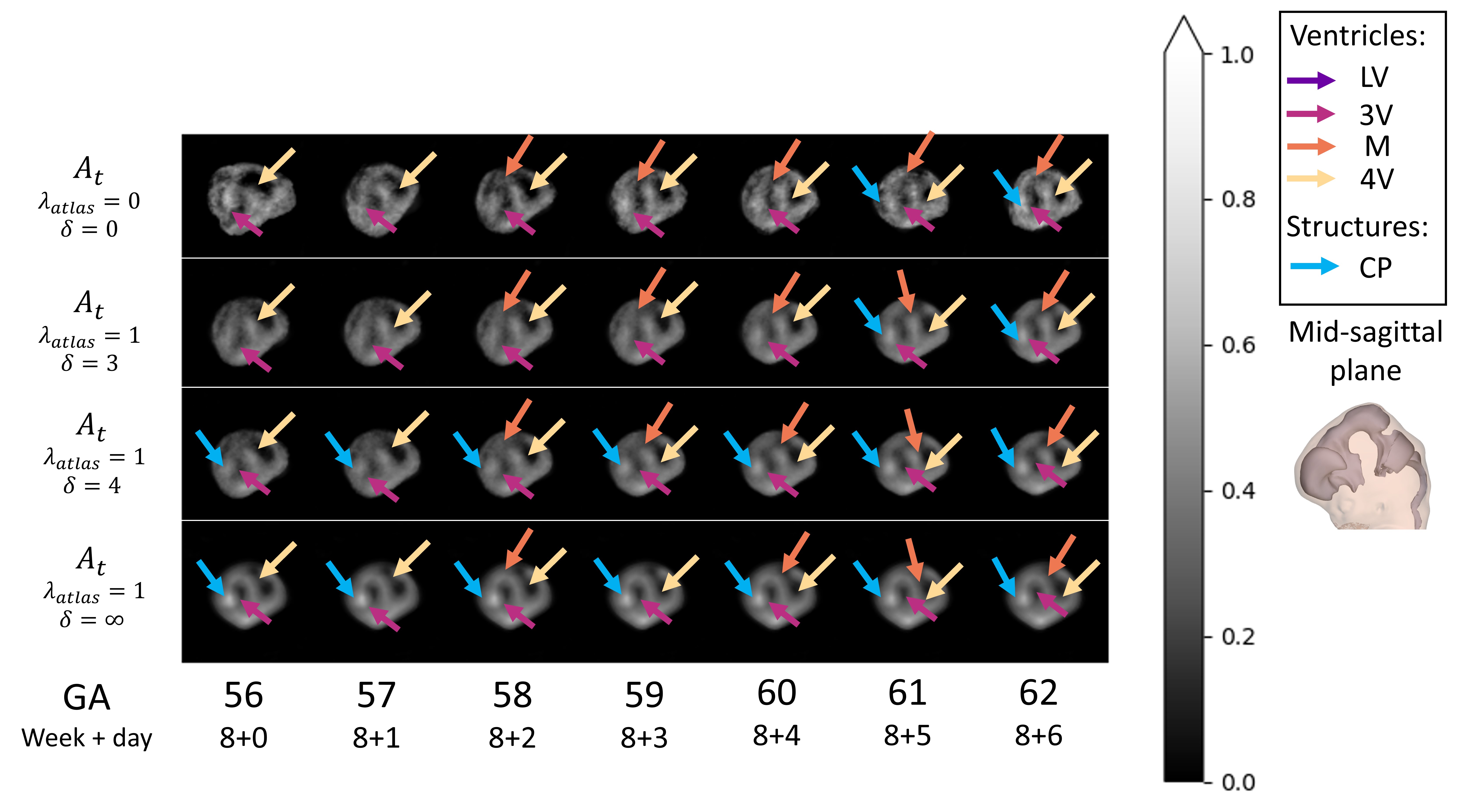}
    \caption{Atlas $A_t$ in gestational week 8 for different values of $\delta$. For $\delta \geq 4$ the choroid plexus appears starting at day 56, which is anatomically incorrect. All images are shown in the mid-sagittal plane. The arrows indicate relevant visible brain structures. LV = lateral ventricle, 3V
= third ventricle (cavity of the diencephalon), M = cavity of the mesencephalon, 4V = fourth ventricle (cavity of the
rhombencephalon), CP = choroid plexus, CER = cerebellum.}
    \label{fig:figure_supp_2}
\end{figure}

\newpage

\section{Supplementary material 2: Subgroup analysis experiment 4}
\begin{figure}[h!]
    \centering
    \includegraphics[width=\linewidth]{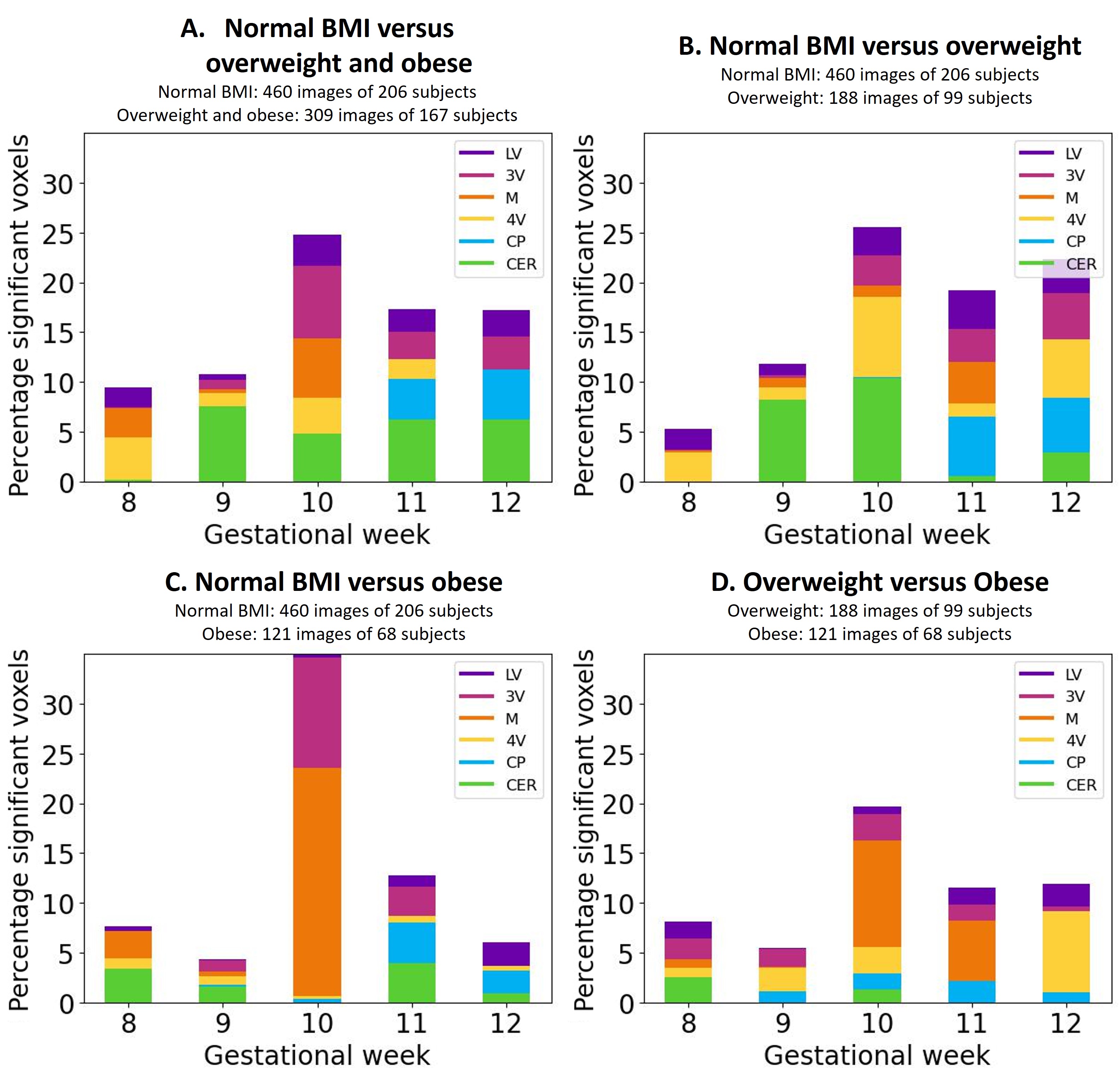}
    \caption{Percentage of significantly different voxels within the delineated ventricles and brain structures at different gestational weeks for the VBM subgroup analysis in Experiment 4. LV = lateral ventricle, 3V = third ventricle (cavity of the diencephalon), M = cavity of the mesencephalon, 4V = fourth ventricle (cavity of the rhombencephalon), CP = choroid plexus, CER = cerebellum.}
    \label{fig:figure_supp_3}
\end{figure}

\begin{figure}[h!]
    \centering
    \includegraphics[width=0.8\linewidth]{figure_16_Supp_fig_3.jpg}
    \caption{Visualization of the comparison between brain morphology for every subgroup analysis using VBM in Experiment 4. The red voxels indicate the p-value for significant structural volumetric differences (p<0.027) and the black arrows highlight the prominent differences between the subgroups. The atlas of the middle of every gestational week is shown. LV = lateral ventricle, 3V = third ventricle (cavity of the diencephalon), M = cavity of the mesencephalon, 4V = fourth ventricle (cavity of the rhombencephalon), CP = choroid plexus, CER = cerebellum.}
    \label{fig:figure_supp_4}
\end{figure}
\end{document}